%% file: sigmod-archived.tex
\documentclass[10pt,sigconf]{acmart}

\copyrightyear{2020} 
\acmYear{2020} 
\setcopyright{acmcopyright}\acmConference[SIGMOD'20]{Proceedings of the 2020 ACM SIGMOD International Conference on Management of Data}{June 14--19, 2020}{Portland, OR, USA}
\acmBooktitle{Proceedings of the 2020 ACM SIGMOD International Conference on Management of Data (SIGMOD'20), June 14--19, 2020, Portland, OR, USA}
\acmPrice{15.00}
\acmDOI{10.1145/3318464.3389757}
\acmISBN{978-1-4503-6735-6/20/06}

\usepackage{multicol}
\usepackage{booktabs} 
\usepackage[ruled,vlined]{algorithm2e}
\usepackage{amsmath}
\usepackage{wrapfig} 
\usepackage{balance}
\usepackage{enumitem}
\usepackage{moresize}
\usepackage{subcaption}
\usepackage{epstopdf}
\usepackage{graphicx} 
\usepackage{marginnote}
\usepackage{color}
\usepackage{subfiles}

\usepackage{multirow} 

\usepackage{bm}
\usepackage{balance}
\usepackage{soul}

\definecolor{myGreen}{rgb}{0.31, 0.78, 0.47}
\definecolor{myRed}{rgb}{0.73, 0.31, 0.28}
\definecolor{myBlue}{rgb}{0, 0.44, 1}
\definecolor{myGrey}{rgb}{0.57, 0.64, 0.69}

\usepackage{array}
\newcommand{\PreserveBackslash}[1]{\let\temp=\\#1\let\\=\temp}
\newcolumntype{C}[1]{>{\PreserveBackslash\centering}p{#1}}
\newcolumntype{R}[1]{>{\PreserveBackslash\raggedleft}p{#1}}
\newcolumntype{L}[1]{>{\PreserveBackslash\raggedright}p{#1}}

\newcommand\Paragraph[1]{\vspace{0.02in}  \noindent \textbf{#1.}}
\newcommand\Paragraphq[1]{\vspace{0.02in}  \noindent \textbf{#1?}}
\newcommand\Paragraphno[1]{\vspace{0.02in}  \noindent \textbf{#1}}
\newcommand\Textcolor[2]{\hspace*{-3mm} \textcolor{#1}{#2} }

\newcommand{\csize}[1]{\texttt{csize(}{#1}\texttt{)}}
\newcommand{\size}[1]{\texttt{size(}{#1}\texttt{)}}

\newcommand{\algo}{\textit{Lethe}}
\newcommand{\comp}{FADE}
\newcommand{\compshort}{FADE}
\newcommand{\layout}{Key Weaving Storage Layout}
\newcommand{\layoutshort}{KiWi}
\newcommand{\dpl}{D_{th}}
\newcommand{\age}[1]{a^{max}_{#1}}

\begin{document}

\title{\algo{}: A Tunable Delete-Aware LSM Engine \\
\huge{(Updated Version)} \vspace*{-1mm}}

\author{Subhadeep Sarkar, Tarikul Islam Papon, Dimitris Staratzis, Manos Athanassoulis}
\affiliation{%
 \institution{Boston University}
}

\begin{abstract}

\input{0-abstract.tex}  
\end{abstract}




\maketitle
\pagestyle{plain} 

\section{Introduction}
\vspace{-0.05in}
\label{sec:intro}
\input{1-intro}

\section{LSM Background}
\label{sec:background}

\input{2-background}

\vspace{-0.05in}
\section{The Impact of Deletes}
\label{sec:design_space}
\input{3-design_space}

\section{Persisting Deletes: \algo}
\label{sec:fade}

\input{4-solution}


\section{Evaluation}
\label{sec:experimental_results}
\input{6-experimental_results}

\section{Related Work}
\vspace{-0.03in}
\label{sec:related_work}

\input{7-related_work}

\section{Conclusion}
\vspace{-0.03in}
\label{sec:conclusion}

\input{8-conclusion}

\balance 
{
\bibliographystyle{abbrv}
\bibliography{library}

} 

\end{document}

%% file: 0-abstract.tex
Data-intensive applications fueled the evolution of 
log structured merge (LSM) based key-value engines that employ 
the \textit{out-of-place} paradigm to support high ingestion rates 
with low read/write interference. 
These benefits, however, come at the cost of \textit{treating
deletes as a second-class citizen}. A delete 
inserts a \textit{tombstone} that invalidates older 
instances of the 
deleted key. State-of-the-art LSM engines do not provide
guarantees as to how fast a tombstone will propagate to 
\textit{persist the deletion}. 
Further, LSM engines only support deletion 
on the sort key. To delete on another attribute 
(e.g., timestamp), the entire tree is read and re-written.
We highlight that fast persistent deletion without affecting
read performance is key to support: (i) streaming
systems operating on a window of data, 
(ii) privacy with latency guarantees on the right-to-be-forgotten, and
(iii) \textit{en masse} cloud deployment of data systems that makes 
storage a precious resource.

To address these challenges, in this paper, we build a new key-value 
storage engine,
\algo{}, that uses a very small amount of additional metadata,
a set of new delete-aware compaction policies, and a new physical 
data layout that weaves the sort and the delete key order.
We show that \algo{} supports any user-defined threshold for the 
delete persistence latency offering \textit{higher read
throughput} ($1.17-1.4\times$) and \textit{lower space amplification} 
($2.1-9.8\times$), with a modest
increase in write amplification (between $4\%$ and $25\%$). In addition,
\algo{} supports efficient range deletes on a 
\textit{secondary delete key} by dropping entire data pages 
without sacrificing read 
performance nor employing a costly full tree merge. 

%% file: 1-intro.tex
\Paragraph{Systems are Optimized for Fast Data Ingestion} 
Modern data systems process an unprecedented amount of data 
generated by a variety of applications that include data analytics, 
stream processing, 
Internet of Things and 5G~\cite{Cisco2018, Gartner2017}. 
Cloud-based latency-sensitive applications 
like live video streaming~\cite{Huang2008}, real-time health 
monitoring~\cite{Pantelopoulos2010},  
e-commerce transactions~\cite{Huang2019}, social network 
analysis~\cite{Sakaki2010}, and online gaming~\cite{Lu2004}, 
generate large volumes of data at a high velocity 
that requires hybrid transactional/analytical processing 
(HTAP)~\cite{Appuswamy2017, Mohan2016, Ozcan2017}. 
Thus, for the past decade, one of the main data 
management research challenges
has been to design data systems that can sustain fast data ingestion 
rate and process queries 
at low latency~\cite{Alagiannis2014, Arulraj2016, Ozcan2017}. 
To achieve this, modern data stores reduce read/write interference by employing \textit{out-of-place} 
ingestion~\cite{Athanassoulis2011, Athanassoulis2015, Dayan2017, Heman2010, Lamb2012, Li2010, Papadopoulos2016, Stonebraker2005}.

\begin{figure}[tb]
    \centering
    \hspace*{-9mm}
    \begin{minipage}{.25\textwidth}
        \centering
        \hspace*{-7mm}
        \includegraphics[scale=0.345]{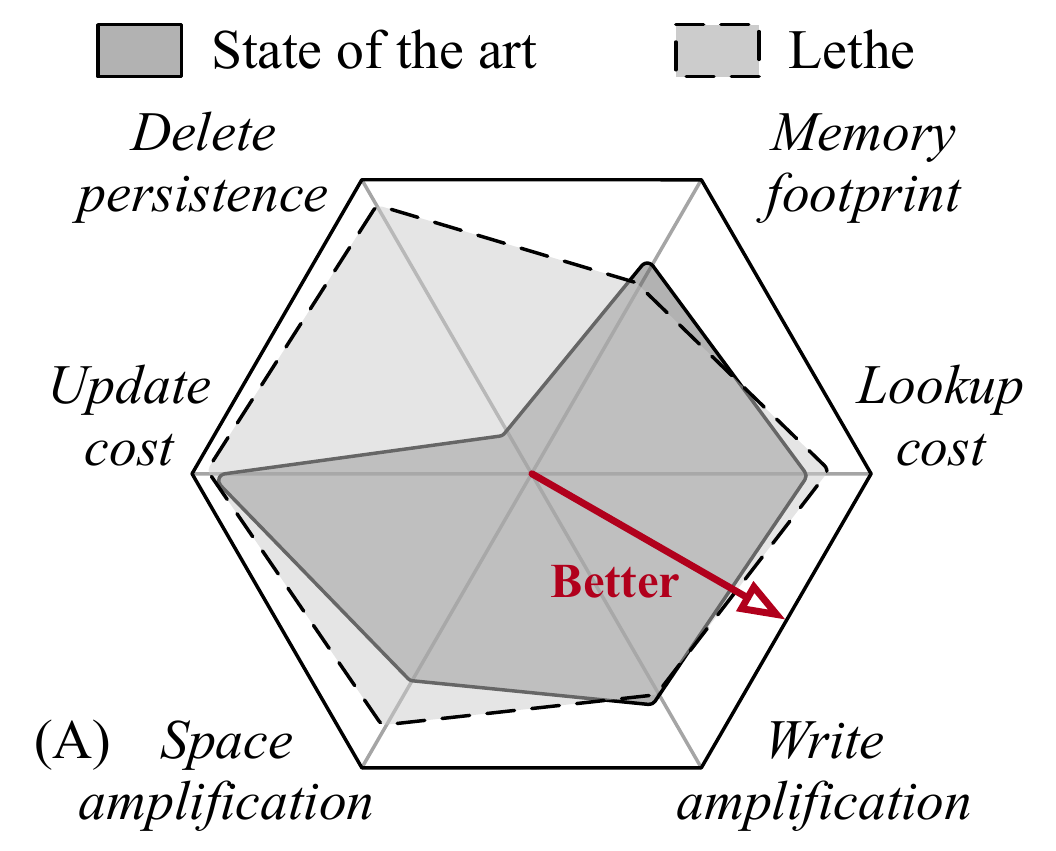}
    \end{minipage}%
    \hspace*{-6mm}
    \begin{minipage}{0.2\textwidth}
        \centering
        \includegraphics[height=1.3in]{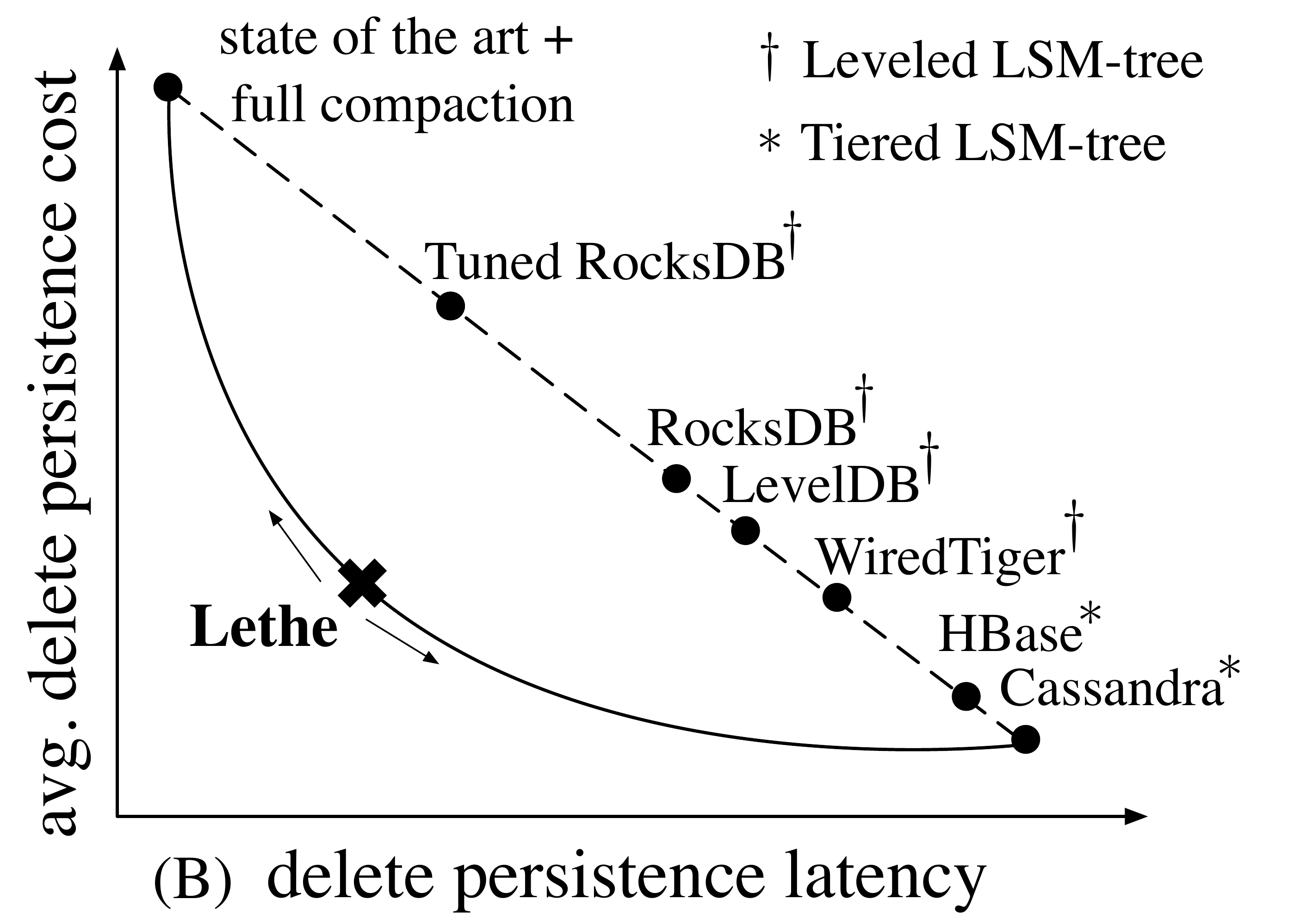}
    \end{minipage}
    \vspace{-0.15in} 
    \caption{(A) \algo{} strikes an optimal balance between the latency/performance for timely delete persistence in LSM-trees,
    and (B) supports timely delete persistence by navigating the latency/cost trade-off.}
     \vspace{-0.25in}
    \label{fig:intro}
\end{figure}

\Paragraph{LSM-based Key-Value Stores} 
The classical \textit{out-of-place} design is the log-structured 
merge (LSM) tree. LSM-trees buffer incoming data
entries in main memory, and periodically flush this buffer as
an \textit{immutable sorted run} on durable 
storage~\cite{Dayan2017, Luo2019, ONeil1996, Raju2017, Thonangi2017}.
In turn, as more sorted runs accumulate, they are
iteratively sort-merged to form fewer yet larger sorted runs. This
process, termed \textit{compaction},
reduces the number of sorted runs accessed during a read query
with amortized merging cost. 
Every compaction sort-merges existing sorted runs from 
consecutive levels and discards any invalid entries. 
LSM-trees are adopted by several modern 
systems including  
LevelDB~\cite{GoogleLevelDB} and BigTable~\cite{Chang2006} at Google, 
RocksDB~\cite{FacebookRocksDB} at Facebook, 
X-Engine~\cite{Huang2019} at Alibaba, 
Voldemort~\cite{LinkedInVoldemort} at LinkedIn, 
Dynamo~\cite{DeCandia2007} at Amazon, Cassandra~\cite{ApacheCassandra}, 
HBase~\cite{ApacheHBase}, and Accumulo~\cite{ApacheAccumulo} at 
Apache, and bLSM~\cite{Sears2012} 
and cLSM~\cite{Golan-Gueta2015} at Yahoo. 
Relational data systems have been increasingly adopting LSM-style
of updates. MyRocks~\cite{FacebookMyRocks} uses RocksDB as storage 
engine and 
SQLite4~\cite{SQLite4} has experimented with LSM-trees in its storage 
layer, while columnar systems use LSM-style 
updates~\cite{Dageville2016,Stonebraker2005,Lamb2012,Zukowski2012}.

\Paragraph{The Challenge: Out-of-place Deletes}
LSM engines use the out-of-place 
paradigm for any write 
operation, including ingestion (inserts), 
modification (updates), and deletion (deletes).
As a result, a delete (update) is implemented by inserting 
additional meta-data that logically invalidates the older target 
entries~\cite{Cao2020}. We refer to this process as \textit{logical deletes 
(updates)}. Both logical deletes and updates show a complex three-way trade-off \cite{Athanassoulis2016}, however, logical deletes have wider 
implications in terms of (i) space amplification, (ii) read cost, 
(iii) write amplification, and (iv) privacy considerations, and 
hence, is the primary focus of this work.

In particular, a logical delete, inserts a \textit{tombstone}
that invalidates all the older entries for a given key, with the
expectation that, \textit{eventually}, they will
be persistently deleted. In practice, the \textit{delete persistence
latency} is 
driven by (a) the system design choices
and (b) workload characteristics. Neither 
can be fully controlled during execution, therefore, providing 
latency guarantees for persistent deletion in state-of-the-art 
LSM engines is nearly impossible. 
In fact, LSM-trees have a potentially unbounded delete 
persistence latency. In order to limit it, current designs 
employ a costly 
full-tree compaction that interferes with read performance and write 
amplification, and results in performance unpredictability \cite{Huang2019}.

\Paragraph{Deletes in LSM-trees} 
LSM-trees are employed as the storage layer for relational 
systems~\cite{FacebookMyRocks}, streaming 
systems~\cite{Akidau2015,Hueske2018,To2017}, and pure 
key-value storage~\cite{MongoDB,WiredTiger}.
As a result, \textbf{an LSM delete operation may be triggered by various logical operations,
not limited to user-driven deletes}. 
For example, deletes are triggered by workloads
that involve \textit{periodic data migration}~\cite{RocksDB2018}, 
streaming operations on a \textit{running window} \cite{Kulkarni2015,Huang2019}, or entail \textit{cleanup during 
data migration} \cite{RocksDB2018}.
In particular, dropping tables in an LSM-tree with multiple column 
families is realized through a range delete operation~\cite{RocksDB2018}.
Another frequent example is data collections sorted on creation 
timestamp. In that case, a classical out-of-place update 
is not enough since the key (timestamp) will also change. Hence, every 
update translates to a delete followed by the insertion of the new 
version \cite{Callaghan2020}.
Below, we distill the common concepts of two frequent delete use-cases.

\vspace{0.02in}
\textit{Scenario 1}: An e-commerce company $EComp$ stores its order 
details sorted by $order\_id$ in an LSM-tree, and needs to 
delete the order history for a particular user.
Within the system, this delete request is translated to a set of 
point and range deletes on the 
\textit{sort key}, i.e., $order\_id$. 

\vspace{0.02in}
\textit{Scenario 2}: A data company $DComp$ stores its operational data in 
an LSM-tree with $document\_id$ as the sort key. As most of the data are 
relevant only for $D$ days, $DComp$ wants to delete all data with a 
$timestamp$ that is older than $D$ days (and archive them). At the same 
time, $DComp$ frequently accesses the documents using $document\_id$, 
hence, the sort key ($document\_id$) is different from the delete 
key ($timestamp$).

\Paragraphq{Why State of the Art is not Enough}
LSM engines cannot
efficiently support $EComp$ from the first scenario for two reasons. First, as deletes 
insert tombstones (retaining the 
physical entries), they \textbf{increase space amplification}. 
Second, 
retaining superfluous entries in the tree \textbf{adversely affects read performance} 
because read queries have to discard potentially large collections of invalid entries, 
which further ``pollute'' the filter metadata~\cite{Huang2019}, and 
\textbf{increase write amplification} because
invalid entries are repeatedly compacted.
Further, LSM engines are ill-suited for
$DComp$ from the second scenario because they cannot efficiently support a range deletion in a delete key other than the sort key (termed 
\textit{secondary range deletes}). Instead, they employ a 
\textbf{full tree compaction}, which causes an excessive number of 
\textbf{wasteful I/Os} while reading, merging, and re-writing the sorted files 
of the \textit{entire} tree \cite{Huang2019}. 

\textit{Delete persistence latency.} 
In order to be able to report
that a \textit{delete persisted}, the corresponding tombstone
has to reach the last level of the tree through iterative 
compactions to discard all invalidated entries. 
The time elapsed between the insertion of the tombstone 
in the tree and the completion of the last-level compaction is
termed \textit{delete persistence latency}. 
\textbf{LSM logical deletes do not provide
delete persistence latency guarantees}, 
hence $EComp$ cannot offer such guarantees to its users. 
In order to add a hard limit on delete persistence latency, 
current designs employ a costly \textbf{full tree
compaction} as well.

\textit{Deletion as privacy.}
Having unbounded delete persistence latency may lead to a breach of privacy.
For example, it was recently reported that 
Twitter retains user messages years after they have been deleted, 
even after user accounts have been 
deactivated~\cite{Whittaker2019}.
With the new data privacy protection acts like GDPR~\cite{Goddard2017} and
CCPA~\cite{CCPA2018}, 
the end-to-end data lifecycle has new privacy challenges to address~\cite{Deshpande2018, Sarkar2018}. 
With user-rights, such as the \textit{right-to-be-forgotten} coming into 
play, persistent deletion within a fixed threshold  
is critical. 

\begin{center}
\vspace{-0.05in}
\textit{``Full tree compactions should be avoided.''}	
\vspace{-0.05in}
\end{center}

\noindent In our interactions with engineers working on LSM-based 
production systems, we learned that periodic 
deletes of a large fraction of data based on timestamp are very frequent. 
To quote an engineer working on XEngine~\cite{Huang2019},
``\textit{Applications may keep data for different durations (e.g., 7 or 
30 days) for their own purposes. But they all have this 
requirement for deletes every day. For example, they may keep data for 30 days, 
and daily delete data that turned 31-days old, effectively purging 1/30 of the database every day.}'' This deletion is performed
with a full tree compaction.  To further quote the same team,
``\textit{Forcing compactions to set a delete latency threshold, leads to significant increase in compaction frequency, and the observed I/O utilization often peaks. This quickly introduces performance pains.}''
For large data companies, 
deleting 1/7 or 1/30 of their database, accounts for several GBs or TBs that is required to be persistently removed daily. 
The current approach of employing full-tree compactions is suboptimal as it 
(1) causes high latency spikes, and (2) increases write amplification.
The goal of this work is to 
address these challenges while \textbf{retaining the benefits of LSM design}.

\Paragraph{The Solution: \algo}
We propose \algo{}\footnote{\textit{Lethe}, the Greek mythological river of oblivion, signifies efficient deletion.}, a new LSM-based 
key-value store that offers efficient deletes without compromising the benefits of LSM-trees. 
\algo{} pushes the boundary of the traditional LSM design space by adding delete 
persistence as a new design goal, and is able to meet user requirements for delete 
persistence latency. Figures \ref{fig:intro} (A) and (B) show a qualitative comparison between 
state-of-the-art LSM engines~\cite{FacebookRocksDB,GoogleLevelDB,WiredTiger,ApacheHBase,ApacheCassandra} 
and \algo{} with respect to the efficiency and cost of 
timely persistent deletes. \algo{} introduces two new LSM design components: 
\textit{\comp{}} and \textit{\layoutshort{}}. 

\textit{\comp{} } (\textbf{Fa}st \textbf{De}letion) is a new family of compaction strategies that
prioritize files for compaction based on (a) the number of 
invalidated entries contained, (b) the age of the 
oldest tombstone, and (c) the range overlap with other files. 
\comp{} uses this information to decide \textit{when} to trigger a compaction on \textit{which} files, to purge invalid entries within a threshold. 

\textit{\layoutshort{}}  (\textbf{Key Wea}ving Storage Layout) is a new continuum of physical layouts that allows for 
tunable secondary range deletes without causing latency spikes, by
introducing the notion of \textit{delete tiles}. An LSM-tree level consists of several
sorted files that logically form a sorted run. \layoutshort{} augments the design of each file 
with several
delete tiles, each containing several data pages. 
A delete tile is sorted on the secondary (delete) key, while each data page remains internally
sorted on the sort key. Having Bloom filters at the page level, and fence pointers
for both the sort key and the secondary delete key, \layoutshort{} facilitates secondary
range deletes by dropping entire pages from the delete tiles, with a constant factor
increase in false positives. Maintaining the pages sorted on the sort key also means
that once a page is in memory, read queries maintain the same efficiency as the
state of the art. 

Putting everything together, \algo{} is the first LSM engine
to our knowledge that offers efficient deletes while improving read performance, 
supports user-defined delete latency thresholds, and enables practical secondary range 
deletes.

\Paragraph{Contributions}
Our contributions are as follows: 

\vspace{-4pt}
\begin{itemize}[leftmargin=*,labelindent=0mm, itemsep=0.2\baselineskip]
    \item We analyze out-of-place deletes w.r.t. read performance, space and write amplification, 
    and user privacy.
    \item We introduce \comp{}
    that bounds delete persistence 
    latency without hurting
    performance and resource consumption.
    \item We introduce \layout{}, the first layout that supports efficient secondary range
    deletes. 
    \item We present the design of \algo{} that integrates \compshort{} and \layoutshort{} in a 
    state-of-the-art LSM engine and enables fast deletes with a tunable balance 
    between delete persistence latency and the overall performance of the system.
    \item We demonstrate that \algo{} offers delete latency guarantees, having up to $1.4\times$ higher read 
    throughput. The higher read throughput is attributed to the significantly lower space amplification 
    (up to $9.8\times$ for only $10\%$ deletes) because it purges
    invalid entries faster. These benefits come at the cost of $4\%$-$25\%$ higher write 
    amplification.
    \item Finally, we demonstrate that \algo{} is the first LSM engine to support efficient
    secondary range deletes at the expense of increased read cost, and we show how to tune \algo{} to 
    amortize this cost based on the workload.
\end{itemize}

\vspace{-0.15in}

%% file: 2-background.tex
\vspace{-0.05in}
\Paragraph{Basics} 
LSM-trees store key-value pairs, 
where a \textit{key} refers to a unique object identifier, and the data associated with 
it, is referred to as \textit{value}. 
For relational data, the primary key acts as the key, and the 
remaining attributes in a tuple constitute the value. 
As entries are sorted and accessed by the key, we 
refer to it as the \textit{sort key}. 
For an LSM-tree with $L$ levels, we assume that its first level (Level $0$) is 
an in-memory buffer and the remaining levels (Level $1$ to $L - 1$) 
are disk-resident. We adopt notation from the literature~\cite{Dayan2017, Luo2019}. 

\Paragraph{Buffering Inserts and Updates} 
Inserts, updates, or deletes are buffered in memory. A 
delete (update) to a key that exists in the buffer, deletes 
(replaces) the older key in-place, otherwise the delete (update)
remains in memory to invalidate any existing instances of the key
on the disk-resident part of the tree.  
Once the buffer reaches its capacity,
the entries are sorted by key to form an  
\textit{immutable sorted run} and are
flushed to the first disk-level (Level $1$). When a disk-level 
reaches its
capacity, all runs within that level are sort-merged and pushed to 
the next
level. To bound the number of levels in a tree,
runs are arranged in exponentially growing levels on disk.
The capacity of Level $i$ ($i \geq 1$) is greater than that of Level $i - 1$ 
by a factor of $T$, termed the size ratio of the tree. 

\Paragraph{Compaction Policies: Leveling and Tiering} 
Classically, LSM-trees support two merging policies: leveling and tiering. 
In leveling, each level may have at most one run, and every time 
a run in Level $i - 1$ ($i \geq 1$) is moved to Level $i$, 
it is greedily sort-merged 
with the run from Level $i$, if it exists. 
With tiering, every level must accumulate $T$ runs before they are 
sort-merged. 
During a sort-merge (compaction), entries with a matching key are 
consolidated and only the most recent valid entry is 
retained~\cite{Dong2017, ONeil1996}. 
Recently hybrid compaction policies fuse
leveling and tiering in a single tree to strike a balance 
between the read and write throughput based on workload 
specifications~\cite{Dayan2018, Dayan2019}.

\textit{Partial Compaction.} To amortize latency spikes 
from compactions in larger levels, state-of-the-art LSM engines 
organize runs into smaller files, and perform compactions at the 
granularity of files instead of levels~\cite{Dong2017}. 
If Level $i$ grows beyond a threshold, a compaction is 
triggered and one file from Level $i$ is chosen to be 
\textit{partially compacted} with 
files from Level $i + 1$ that have an overlapping key-range.
Deciding which file to compact depends on the storage engine design. 
For instance, to optimize write throughput, we select files from Level $i$ 
with minimal overlap with files in Level $i + 1$, 
to minimize both write amplification and compaction time.

\Paragraph{Querying LSM-Trees}
A point lookup begins at the memory buffer and 
traverses the tree from the smallest disk-level to 
the largest one. For tiering, within a level, 
a lookup moves from the most to the least recent tier. 
The lookup terminates when it finds the first matching entry. 
A range lookup returns the most recent versions of the target keys 
by sort-merging the qualifying key ranges across all runs in the tree.

\Paragraph{Optimizing Lookups} 
Read performance is optimized using Bloom filters (BFs) and fence pointers. 
In the worst case, a lookup 
needs to probe every run. To reduce this cost, 
LSM engines use one BF per run in 
main memory~\cite{Dayan2017, FacebookRocksDB}. 
\textit{Bloom filters} allow a lookup to skip 
probing a run altogether if the filter-lookup returns negative. In 
practice, for efficient storage, BFs are maintained at the 
granularity of files~\cite{Dong2017}. 
\textit{Fence pointers} store the smallest key per disk page in 
memory~\cite{Dayan2017}, to quickly identify which page(s)
to read for a lookup, and perform up to one I/O per run for point 
lookups.

%% file: 3-design_space.tex
We now describe the design space of deletes in LSM-trees.

\vspace{-0.1in}
\subsection{Delete Design Space}
\vspace{-0.05in}
In LSM-trees an entry at Level $i$ 
is always more recent than an entry with the same key at Level $j$, if $j > i$. 
LSM-trees exploit this to logically delete using tombstones that supersede 
older entries with a matching key. The left part of Figure \ref{fig:problem} 
shows a leveled LSM-tree, the structure of a key-value pair, and a tombstone. 
A key-value contains typically many attributes as part of the value, and a tombstone 
consists of the (deleted) key and the tombstone flag.
  
\vspace{-0.05in}
\subsubsection{\textbf{Primary Deletes}} 
We discuss deletes on the sort key.

\begin{figure}[t]
    \centering
    \includegraphics[scale=0.19]{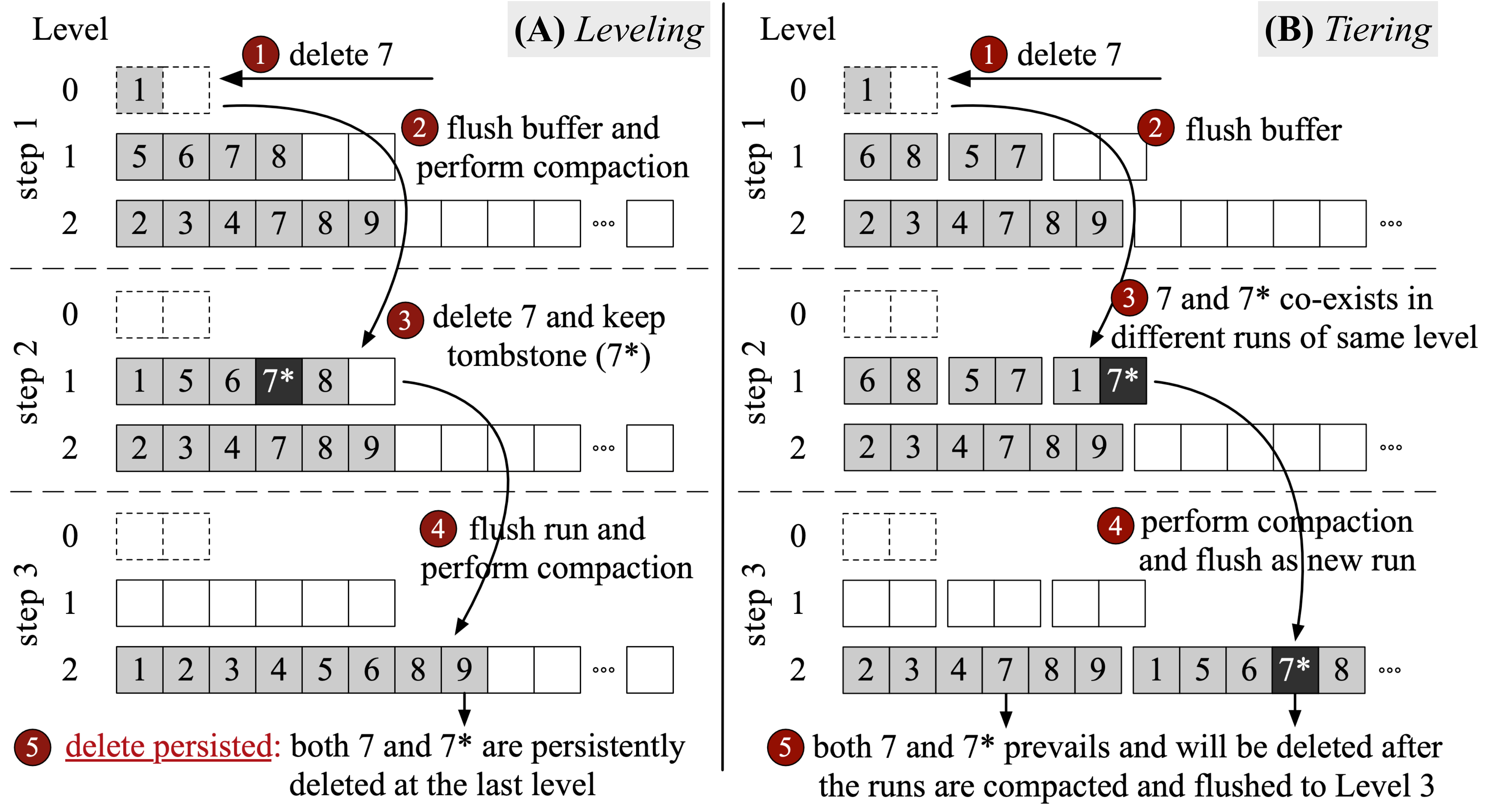}        
    \vspace{-0.1in}
    \caption{In an LSM-tree, for every tombstone, there can be 
    \textbf{(A)} one matching entry per level for leveling or 
    \textbf{(B)} one matching entry per tier per level ($T$ per level) for tiering, where $T=3$ in this example. }
    \label{fig:deletes_LSM} 
    \vspace{-0.25in}
\end{figure}

\Paragraphno{Point Deletes}
insert a tombstone against the key to be deleted 
(Figure \ref{fig:deletes_LSM}). 
Within memory buffer, the tombstone 
replaces in-place any older entry with a matching key. 
On disk, the tombstones are stored within a run in sorted order 
along with other key-value pairs. During compaction, 
a tombstone deletes older entries with the same key and is retained
as there might be more (older) entries with the 
same delete key in subsequent compactions (Fig. \ref{fig:deletes_LSM}).
A tombstone is discarded during its compaction with the last 
level of the tree,  
making the logical delete \textit{persistent}.

\Paragraphno{Range Deletes} in LSM-trees are common, 
however, they cause performance problems. 
Range deletes generate special \textit{range 
tombstones} that are stored in a separate \textit{range tombstone block} within 
files~\cite{FacebookRocksDB}. During data access, a histogram
storing deleted ranges is maintained in memory which has to
be accessed by every point query slowing down read 
accesses~\cite{Callaghan2020,RocksDB2018}.
Similar to point deletes, range deletes are persisted 
when the files that contain them are compacted with the 
last level, leaving potentially unbounded persistence latency. 
Thus, in practice, a complete full tree compaction is periodically 
employed to ensure delete persistence~\cite{Huang2019}. 
During such compactions, all reads and writes to the data store 
are stalled, which results in latency spikes.

\begin{figure*}[!ht]
    \centering  
        \includegraphics[scale=.175]{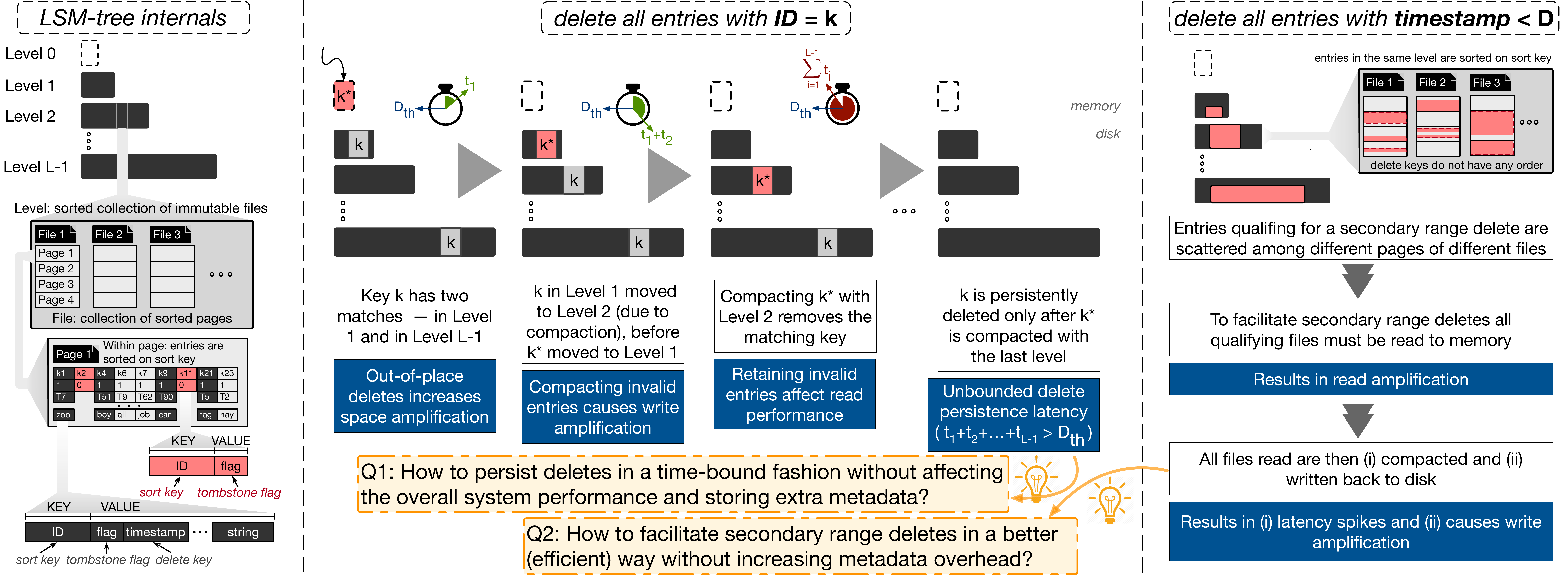}  
        \vspace*{-2mm}        
    \caption{The design space of deletes in LSM-trees.}
    \label{fig:problem} 
\vspace*{-0.15in}        
\end{figure*}

\Paragraph{Persistence Latency} 
The latency for persisting a logical delete depends on the 
workload and the data size. 
The middle part of Figure \ref{fig:problem} illustrates 
the operation ``\textit{delete all entries with} 
\texttt{ID} = \texttt{k}''.
Within the system, the operation  
inserts
a tombstone, $\texttt{k}^*$, that logically \textit{invalidates} 
$\texttt{k}$. 
On disk, entries with key $\texttt{k}$ may be located at any level between $1$ and 
$L$. Thus, to ensure delete persistence, $\texttt{k}^*$ must 
participate in $L$ compactions, one at each level of the tree. 
Since compactions are triggered when a level reaches a nominal 
capacity, the \textit{rate of unique insertions} is 
effectively driving the compactions. 
The size of a level grows
exponentially with $T$, therefore, a taller tree
requires exponentially more unique insertions to propagate a 
tombstone to the last level. Hence, the delete persistence 
latency depends on (i) the rate of unique insertions and 
(ii) the current height of a tree.

\Paragraph{Adversarial Workloads} 
Tombstones may be recycled in intermediate 
levels of the tree leading to
\textit{unbounded delete persistence latency} and 
perpetual retention of invalid entries~\cite{Callaghan2020}. 
For example, a workload that 
mostly modifies hot data (in the first few levels)
will grow the tree very slowly, keeping its structure
mostly static. 
Another example is a workload with interleaved inserts and deletes, with 
the deletes issued on a few recently inserted entries that are  
at the smaller levels. In both cases, a newly inserted tombstone
may be recycled in compactions high up the tree that  consolidate entries
rather than propagate towards the last level.

\vspace{-0.05in}
\subsubsection{\textbf{Secondary Deletes}}
We refer to deletes based on an attribute other than
the sort key as \textit{secondary deletes}. The most common type of
secondary deletes is a \textbf{secondary range delete}. 
Consider the operation
``\textit{delete all entries that are older than $D$ days}'',
similar to the second scenario from the introduction.
In the left part of Figure \ref{fig:problem}, we highlight  
the sort key ($\texttt{ID}$) and the delete key ($\texttt{timestamp}$) 
of a key-value pair.
As the entries in a tree are sorted on the sort key, an entry with 
a qualifying delete key may be anywhere in the tree, and this delete pattern
is not efficiently supported. 

\vspace{-0.1in}
\subsubsection{\textbf{Limitations of the State of the Art}} 
In state-of-the-art LSM engines, deletes are considered as 
``second-class citizens''. 
In practice, to ensure time-bounded persistence of logical deletes and to 
facilitate secondary range deletes, 
data stores resort to \textit{periodic full-tree 
compaction}~\cite{Corbett2012, Huang2019}. 
However, this is an extremely expensive solution as 
it involves superfluous disk I/Os, increases write amplification and results in 
latency spikes. 
To reduce excessive I/Os, RocksDB implements a file selection 
policy based on the number of 
tombstones~\cite{FacebookRocksDB}. This reduces the amount of invalid
entries, but it does not offer persistent delete latency guarantees.

\vspace{-0.1in}
\subsection{Implications of Out-of-place Deletes}
Next, we quantify the implications of out-of-place deletes on
read performance, and space and write amplification.

\Paragraph{Model Details} 
We assume an LSM-tree with size ratio $T$, that stores $N$ entries across
$L+1$ levels. The size of the memory buffer 
is $M = P \cdot B \cdot E$, where $P$ is the number of disk
pages in the buffer, $B$ is the number of entries per page, 
and $E$ is the average size of an entry. The capacity
of this tree is $\sum_{i=0}^L M \cdot T^i$, 
where $M \cdot T^i$ is the 
capacity of Level $i$. The $N$ entries inserted in the tree includes $\delta_p$ 
point tombstones and $\delta_r$ range tombstones that have an average selective 
of $\sigma$. Table \ref{tab:notation} shows all the 
parameters used in our modeling.

\subsubsection{\textbf{Space Amplification}}
\label{subsubsec:SA}
Deletes increase space amplification by 
(i) the tombstones and (ii) the invalidated entries (for every key,
there might be several invalid versions).
Space amplification increases storage cost and the overhead for data 
organization (sorting) and processing (read I/Os during compaction). 
Commercial databases often report space amplification of about 
$11\%$~\cite{RocksDB2018}, however, this corresponds to $T = 10$, a single point 
in the vast design continuum.

\Paragraph{Analysis} 
Following prior work~\cite{Dayan2018}, we define space amplification 
as the ratio between the size of superfluous 
entries and the size of the unique entries in the tree, 
$s_{amp} = \tfrac{\csize{N} - \csize{U}}{\csize{U}}$, 
where $\csize{N}$ is the cumulative size of all entries 
and $\csize{U}$ is the cumulative size of all unique entries. 
Note that $s_{amp}\in [0,\infty)$, and that if all inserted 
keys are unique there is no space amplification.

\Paragraph{Without Deletes} 
Assume a workload with inserts and updates (but no deletes) 
for a leveled LSM-tree. In the worst case, all entries in levels up to $L - 1$ can be 
updates for the entries in Level $L$, 
leading to space amplification $O(1/T)$. 
For a tiered LSM-tree, the worst case is when the tiers of a level overlap, and the first $L-1$ levels contain updates 
for Level $L$. 
This leads to space amplification $O(T)$. 

\Paragraph{With Deletes} 
If the size of a tombstone is the same as the size of a key-value entry, 
the asymptotic worst-case space 
amplification remains the same as that with updates for leveling.  
However, in practice, a tombstone is orders of magnitude smaller than 
a key-value entry. We introduce the tombstone size ratio
{\ssmall$\lambda = \tfrac{\size{tombstone}}{\size{key\text{-}value}} 
\approx \tfrac{\size{key}}{\size{key} + \size{value}}$}, 
where $\size{key}$ and $\size{value}$ is the average size of a key and an entry, 
respectively. $\lambda$ is bounded by $(0, 1]$, and a smaller $\lambda$ implies 
that a few bytes (for tombstones) can invalidate more bytes 
(for key-values) and lead to larger space amplification given by 
{\small$O\left(\frac{(1-\lambda) \cdot N + 1}{\lambda \cdot T}\right)$}. 
For tiering, in the worst case, tombstones in the recent-most tier 
can invalidate all entries in that level, resulting in 
space amplification $O\left(\frac{N}{1 - \lambda}\right)$.

\vspace{-0.1in}
\subsubsection{\textbf{Read Performance}} Point tombstones are hashed to the BFs 
the same way as valid keys, and thus, increase the false positive 
rate (FPR) for the filters as well as the I/O cost for point lookups. 
Also, deleted entries cause range queries to scan invalid data before finding 
qualifying keys. 
Consider that a range delete with $0.5\%$ selectivity over a 100GB database 
invalidates 500MB, which might have to be scanned (and discarded)
during query execution.

\Paragraph{Analysis: Point Lookups} 
A point lookup probes one (or more) BF before performing any disk I/O.
The FPR of a BF depends on the number of bits allocated to the  
filter in the memory ($m$) and the number of entries ($N$) hashed into 
the filter, 
and is given by $e^{- m/N \cdot (ln(2))^2  }$. 
For leveling, the average worst-case point lookup cost on non-existing entries 
is $O(e^{-m/N})$, and for tiering, the cost becomes 
$O(T \cdot e^{-m/N})$~\cite{Dayan2017}. 
For lookups on existing entries, this 
cost increases by $1$ as the lookup has to probe at least one page.
Since tombstones are hashed into the BFs, retaining tombstones 
and invalid entries increases their FPR, thus hurting point lookups.

\Paragraph{Analysis: Range Lookups} 
A range query on the sort key reads and merges all qualifying disk pages.
The I/O cost of a \textit{short range query} accessing one page per level 
is $O(L)$ for leveling and $O(L \cdot T)$ for tiering. The I/O cost for 
long range lookups depends on the selectivity of the lookup range, and is 
$O(s \cdot N/B)$ for leveling and $O(s \cdot T  \cdot N/B)$ for tiering. 
When answering range queries, tombstones and invalid entries have to be read
and discarded, slowing down the range queries.

\vspace{-0.1in}
\subsubsection{\textbf{Write Amplification}}
\label{subsubsec:WA}
Before being consolidated, an invalid entry may participate 
in multiple compactions. Repeatedly compacting invalid
entries increases write amplification,
which is particularly undesirable for installations that the durable
storage has limited write endurance~\cite{RocksDB2018}. 

\Paragraph{Analysis} We define write amplification, $w_{amp}$ as the ratio of the 
total bytes written on disk that correspond to unmodified entries to the total 
bytes written corresponding to new or modified entries, 
$w_{amp} = \tfrac{\csize{N^+} - \csize{N}}{\csize{N}}$. $N^+$ is the number of
all the entries written to disk including the entries re-written as unmodified after 
a compaction. For leveling, every entry participates on average in $T/2$ compactions 
per level which makes $N^+ = N \cdot L \cdot T/2$. For tiering, every entry is  
written on disk once per level, implying $N^+ = N \cdot L$. 
Thus, $w_{amp}$ for leveled and tiered LSM-trees are given by 
$O(L \cdot T)$ and $O(T)$, respectively. 
Note that, as the data size increases, entries participate 
in more compactions unmodified including invalid entries
further increasing write amplification.

\vspace{-0.1in}
\subsubsection{\textbf{Persistence Latency and Data Privacy}} 
The lack of guarantees in persistence latency
has severe implications on data privacy.
With new data privacy protection acts~\cite{CCPA2018,Goddard2017} and the increased 
protection of rights like the 
\textit{right-to-be-forgotten}, data companies are legally obliged to persistently 
delete data offering guarantees~\cite{Deshpande2018} 
and rethink the end-to-end data 
lifecycle~\cite{Sarkar2018}.

\begin{table}[t]
{
\ssmall
    \centering
    \resizebox{0.475\textwidth}{!}{%
    \begin{tabular}{|C{4.5mm}|l|l|}
    \hline
    \textbf{\hspace*{-1mm}Sym.}   & \textbf{Description}    & \textbf{Reference value}   \\
    \hline
    \multirow{1}{*}{$N$}    & \multirow{1}{*}{\begin{tabular}[c]{@{}l@{}}  \# of entries inserted in tree (including tombstones) \end{tabular}}        &   \multirow{1}{*}{$2^{20}$ entries}     \\ 
        
    \hline
    \multirow{1}{*}{$T$}    & \multirow{1}{*}{size ratio of tree}        &   $10$     \\
    \hline
    \multirow{1}{*}{$L$}    & \multirow{1}{*}{number of tree-levels on disk with $N$ entries}         &   \multirow{1}{*}{3 levels}     \\ 
    \hline
    \multirow{1}{*}{$P$}    & \multirow{1}{*}{size of memory buffer in disk pages}        &   $512$ disk pages     \\
    \hline
    \multirow{1}{*}{$B$}    & \multirow{1}{*}{number of entries in a disk page}        &   $4$ entries     \\
    \hline
    \multirow{1}{*}{$E$}    & \multirow{1}{*}{average size of a key-value entry}        &   $1024$ bytes     \\
    \hline
    \multirow{1}{*}{$M$}    & \multirow{1}{*}{memory buffer size }        &   $16$ MB     \\
    \hline
    \multirow{1}{*}{$\delta_p$}    & \multirow{1}{*}{number of point deletes issued}        &  $3 \times 10^5$ entries     \\ 
    \hline
    \multirow{1}{*}{$\delta_r$}    & \multirow{1}{*}{number of range deletes issued}        &    $10^3$ entries  \\ 
    \hline
    \multirow{1}{*}{$\sigma$}    & \multirow{1}{*}{average selectivity of range deletes}        &  $5 \times 10^{-4}$     \\ 
    \hline
    \multirow{1}{*}{$N_\delta$}    & \multirow{1}{*}{\begin{tabular}[c]{@{}l@{}} approx. \# of entries after persisting deletes \end{tabular}}        &   \multirow{1}{*}{-}     \\ 
    \hline
    \multirow{1}{*}{$\lambda$}    & \multirow{1}{*}{\begin{tabular}[c]{@{}l@{}} tombstone size / average key-value size \end{tabular}}        &     \multirow{1}{*}{$0.1$}    \\ 
    \hline
    \multirow{1}{*}{$I$}    & \multirow{1}{*}{ingestion rate of unique entries in tree}        &   $1024$ entries/sec     \\
    \hline
    \multirow{1}{*}{$s$}    & \multirow{1}{*}{selectivity of a long range lookup}        &     -   \\
    \hline
    \multirow{1}{*}{$L_\delta$}    & \multirow{1}{*}{number of tree-levels on disk with $N_\delta$ entries}        &   -     \\ 
    \hline
    \multirow{1}{*}{$m$}    & \multirow{1}{*}{total main memory allocated to BFs}        & $10$  MB     \\
    \hline
    \multirow{1}{*}{$h$}    & \multirow{1}{*}{number of disk pages per delete tile}        &   $16$ disk pages     \\
    \hline
    \end{tabular}
    }
}   
\caption{\algo{} parameters.  \label{tab:notation}}
\vspace{-0.35in}
\end{table}

\Paragraph{Analysis} 
We define \textit{delete persistence latency} as the 
worst-case 
time required, following the insertion of a tombstone, to ensure that the tree is
void of any entry with a matching (older) key to that of the tombstone.
This time depends on 
the insertion rate of unique key-value
entries ($I$) and the height of the tree ($L-1$), and is the time needed to
insert the minimum number of unique keys that is sufficient to
trigger enough compactions. For leveling, delete
persistence latency is $O\left( \frac{T^{L-1} \cdot P \cdot B}{I} \right)$ and for 
tiering is $O\left( \frac{T^L \cdot P \cdot B}{I} \right)$.
This shows that for an LSM-tree with large number of entries ($T^L$) that is built by 
an update-intensive workload, 
the delete persistence latency can be remarkably high. 

\vspace*{-2mm}
\subsection{Implications of the Storage Layout} 
Every file of an LSM-tree is sorted using the sort key.
While this supports read, update, and delete queries on the sort key it cannot support 
operations on a secondary attribute.

\Paragraphno{Secondary Range Deletes} on a delete key that is different from 
the sort key can only be supported by eagerly performing a full tree compaction,
because there is no way to identify the affected files. This stalls all write 
operations, causing huge latency spikes. 
The cost incurred by a secondary range delete depends on the total number of data 
pages on disk, and is independent of the selectivity of the range delete 
operation. Irrespective of the merging strategy, this cost is $O(N/B)$, where $B$ is the 
page size.

%% file: 4-solution.tex
\vspace{-0.05in}
\Paragraph{Design Goals} \algo{} aims (i) to provide persistence guarantees for point 
and range deletes and (ii) to enable practical secondary range deletes. We achieve the 
first design goal by introducing \comp{}, a family of delete-aware compaction strategies. 
We achieve the second goal by introducing \layout{}, a new continuum of physical
data layouts that arranges entries on disk in an interweaved fashion based on both
the sort and the delete key.

\begin{figure*}[!ht]
    \centering  
        \includegraphics[scale=.225]{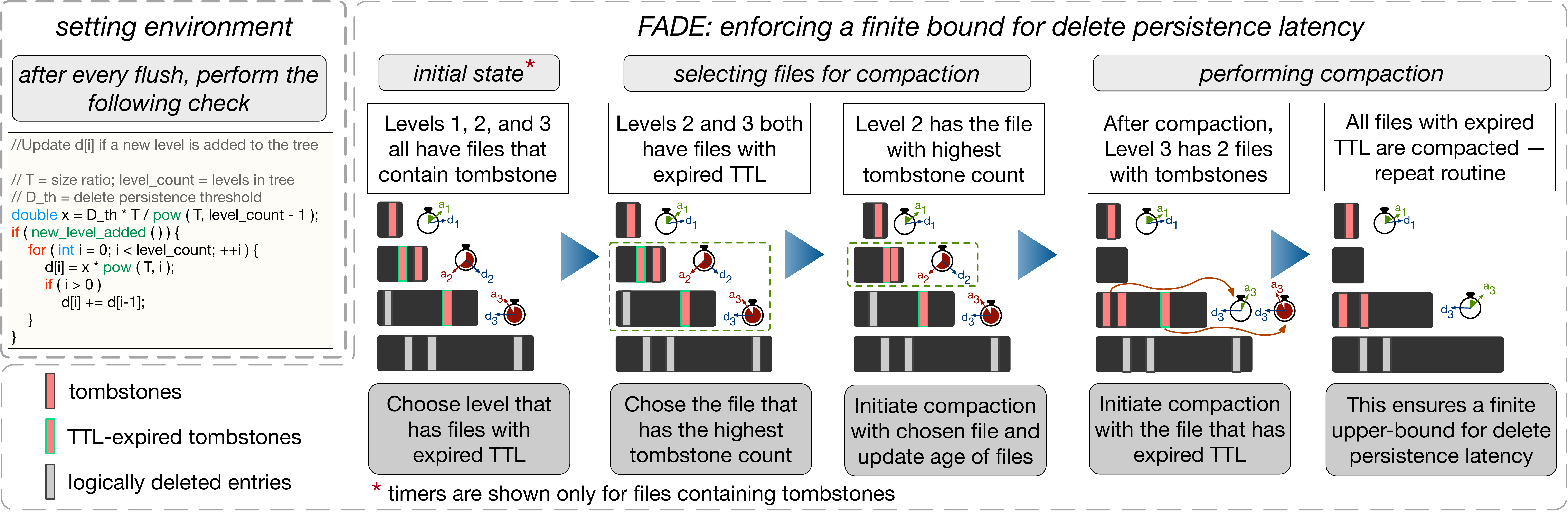} 
        \vspace*{-2mm}        
    \caption{\comp{} persists tombstones within the delete persistence threshold, thus, improving overall performance.}
    \label{fig:ec} 
        \vspace*{-0.15in}        
\end{figure*} 

\vspace*{-2mm}
\subsection{\comp}
We first introduce the \textit{\comp} family of compaction strategies that
ensures that all tombstones are persisted 
within a delete persistence threshold ($\dpl$). 
$\dpl$ is typically specified by the application or user~\cite{Deshpande2018, Sarkar2018} 
as part of the 
service level agreement (SLA) that concerns the data retention policy. 
All data streams bound by the same data retention SLA, 
have the same delete persistence latency.

\vspace{-0.05in}
\subsubsection{\textbf{Overview}} 
Compactions in LSM-trees influence their behavior by dictating
their space amplification, write amplification,
point and range read performance, and delete persistence latency. 
\comp{} uses additional information about the \textit{age of a file's
tombstones} and the \textit{estimated invalidated entries per tombstone}
to ensure that every tombstone will adhere to the user/application-provided 
$\dpl$ by assigning to every file
a \textit{time-to-live} (TTL). As long as $\dpl$
is respected, \comp{} offers different strategies for secondary optimization
goals including minimizing write amplification, minimizing space amplification,
or maximizing system throughput.

\vspace{-0.05in}
\subsubsection{\textbf{Time-to-Live}}
To ensure that all delete operations issued on an LSM-tree 
are persisted before $\dpl$, \compshort{} propagates each tombstone through all 
intermediate levels to the last level within that threshold from its insertion.
\compshort{} achieves this by assigning a smaller TTL
for every file in every level $d_i$, such that $\sum_{i=0}^{L-1} d_i = \dpl$.
A simple TTL allocation is to use $d_i=\dpl/L$.
While this may guarantee that a tombstone reaches the last level within $\dpl$, it
also leads to increased compaction time and resource starvation
as larger levels have exponentially more files, hence, a large number of files 
may exhaust their TTL simultaneously. 
Instead, we assign exponentially increasing TTL per level to guarantee
that files expire at a constant rate per time unit.

\Paragraph{Computing \pmb{$d_i$}} 
For a tree with size ratio $T$, we set the
TTL for level $i$ to be $d_i = T \cdot d_{i-1}$, 
$\forall i \in \{1, L-1\}$ and $d_0 = \dpl \cdot \tfrac{T - 1}{T^{L - 1} - 1}$. Note 
that both the TTL and the capacity per level increase exponentially
for larger levels of the tree.

\Paragraph{Updating \pmb{$d_i$}} 
For a given tree height every file is assigned
a TTL depending on the level it is being compacted into. As more data entries are
inserted, the tree might grow in height. At that point, the TTLs throughout the tree 
have to be updated.
The cost of calculating $d_i$ is low, hence, \compshort{} re-calculates $d_i$ after
every buffer flush. Step $1$ in Figure \ref{fig:ec} (A) shows how to 
update $d_i$ when a new level is added.

\vspace{-0.05in}
\subsubsection{\textbf{\compshort{} Metadata}}
Tombstones for point deletes are stored along with valid key-value pairs, and 
range tombstones are stored in a separate block. 
In addition to the tombstones, \compshort{} requires the values of two 
metrics per file: (i) the \textit{age} of the oldest tombstone contained ($\age{}$) 
and (ii) the \textit{estimated invalidation count} ($b$) of the file tombstones.
After every flush, a file is assigned with its current $\age{}$ and $b$.

In practice, LSM-engines store file metadata including (i) the file creation 
timestamp, and (ii) the distribution of the entries per file in the form of a histogram.
For example, RocksDB assigns a monotonically increasing insertion-driven 
\textit{sequence number} (\texttt{seqnum}) to all entries, and stores the 
number of entries (\texttt{num\_entries}) and point tombstones (\texttt{num\_deletes}) 
for every file. \compshort{} takes advantage of this existing metadata. It uses 
\texttt{seqnum} to compute $\age{}$ and uses 
\texttt{num\_entries} and \texttt{num\_deletes} to compute $b$. Thus, 
in practice \compshort{} leaves no metadata footprint.

\Paragraph{Computing \boldsymbol{$\age{}$}}
The $\age{}$ of a file $f$, termed $\age{f}$, is the age of the oldest (point or range) 
tombstone contained in a file, and is calculated using the difference between the 
current system time and time the oldest tombstone of that file was inserted in the
memory buffer. File without tombstones have $\age{f}=0$. Storing $\age{f}$ requires 
one timestamp (8 bytes) per file, a negligible overhead.

\Paragraph{Computing \pmb{$b$}}
The estimated number of invalidated entries by the tombstones of a file $f$,
termed $b_f$, is calculated using (i) the exact count of point tombstones in
the file ($p_f$) and (ii) an estimation of the entries of the entire database
invalidated by the range tombstones of the file ($rd_f$), as $b_f=p_f+rd_f$.
It is not possible to accurately calculate $rd_f$ without accessing the entire
database, hence, we \textit{estimate} this value using the system-wide histograms
that are already maintained by the data store. The value of $b_f$ is computed 
on-the-fly without needing any additional metadata.

\Paragraph{Updating \boldsymbol{$\age{}$} and \pmb{$b$}} 
Similarly to all file metadata, $\age{}$ and $b$ are first computed 
when a file is created after a buffer flush. Thereafter, for newly 
compacted files, $\age{}$ and $b$ are recomputed before they are written 
back to disk. When a compaction simply moves a file 
from one disk level to the next without
physical sort-merging (i.e., when there are no overlapping keys), $b$ 
remains unchanged and $\age{}$ is recalculated based on the time of the 
latest compaction. 
Note that since all metadata is in memory, this does not cause an I/O. 

\vspace{-0.05in}
\subsubsection{\textbf{Compaction Policies}}
Compactions ensure that both insert and read costs are amortized.
For every compaction, there are two policies to be decided: the \textit{compaction 
trigger policy} and the \textit{file selection policy}. State-of-the-art LSM engines
initiate a compaction when a level is saturated (i.e., larger than a nominal size 
threshold) and either pick a file at random, or the one with the smallest overlap 
with the subsequent level to minimize the merging cost.

\Paragraph{Compaction Trigger}
\compshort{} augments the state-of-the-art by triggering a compaction, not only when
a level is saturated, but also when a file has an expired TTL. \comp{} 
triggers a compaction in a level that has at least one file with expired TTL regardless of its saturation. 
If no TTL has expired, but a 
level is saturated, a compaction in that level is triggered.

\begin{table*}[t]
    \centering
    \resizebox{\textwidth}{!}{%
        \begin{tabular}{l|cc|cC{0mm}cC{0mm}|cC{0mm}cC{0mm}|cC{0mm}cC{0mm}}
        \toprule
        \multicolumn{1}{c}{\multirow{2}{*}{\begin{tabular}[c]{@{}c@{}}\textbf{Metric} \end{tabular}}}  & \multicolumn{2}{c}{\textbf{State-of-the-art~\cite{Dayan2017, Dayan2018}}}                     & \multicolumn{4}{c}{\textbf{\comp}}                     & \multicolumn{4}{c}{\textbf{\layout}}                   & \multicolumn{4}{c}{\textbf{\algo}} \\
        \multicolumn{1}{c}{}                                                                    & \multicolumn{1}{c}{Leveling}        & \multicolumn{1}{c}{Tiering}   & \multicolumn{2}{c}{Leveling}        & \multicolumn{2}{c}{Tiering}    & \multicolumn{2}{c}{Leveling}        & \multicolumn{2}{c}{Tiering}    & \multicolumn{2}{c}{Leveling}        & \multicolumn{2}{c}{Tiering}    \\
        \midrule
        \multirow{1}{*}{Entries in tree}             & $O(N)$          & $O(N)$           & $O(N_\delta)$ & \Textcolor{myGreen}{\boldsymbol{$\blacktriangle$}}   & $O(N_\delta)$  & \Textcolor{myGreen}{$\blacktriangle$}    & $O(N)$ & \Textcolor{myGrey}{\boldsymbol{$\bullet$}}   & $O(N)$  & \Textcolor{myGrey}{\boldsymbol{$\bullet$}}   & $O(N_\delta)$ & \Textcolor{myGreen}{\boldsymbol{$\blacktriangle$}}  & $O(N_\delta)$ & \Textcolor{myGreen}{\boldsymbol{$\blacktriangle$}}   \\ 
        \midrule
        \multirow{1}{*}{Space amplification without deletes}         & $O(1/T)$    & $O(T)$    & $O(1/T)$   & \Textcolor{myGrey}{\boldsymbol{$\bullet$}}    & $O(T)$   & \Textcolor{myGrey}{\boldsymbol{$\bullet$}}  & $O(1/T)$   & \Textcolor{myGrey}{\boldsymbol{$\bullet$}}    & $O(T)$   & \Textcolor{myGrey}{\boldsymbol{$\bullet$}}    & $O(1/T)$   & \Textcolor{myGrey}{\boldsymbol{$\bullet$}}    & $O(T)$   & \Textcolor{myGrey}{\boldsymbol{$\bullet$}} \\ 
        \midrule
        \multirow{1}{*}{Space amplification with deletes}         & $O\left(\frac{(1-\lambda) \cdot N + 1}{\lambda \cdot T}\right)$    & $O\left(\frac{N}{1 - \lambda}\right)$    & $O\left(1/T\right)$  & \Textcolor{myGreen}{\boldsymbol{$\blacktriangle$}}    & $O(T)$  & \Textcolor{myGreen}{\boldsymbol{$\blacktriangle$}}  & $O\left(\frac{(1-\lambda) \cdot N}{\lambda \cdot T}\right)$ & \Textcolor{myGrey}{\boldsymbol{$\bullet$}}   & $O\left(\frac{N}{1 - \lambda}\right)$ & \Textcolor{myGrey}{\boldsymbol{$\bullet$}}   & $O(1/T)$   & \Textcolor{myGreen}{\boldsymbol{$\blacktriangle$}}    & $O(T)$   & \Textcolor{myGreen}{\boldsymbol{$\blacktriangle$}} \\ 
        \midrule
        \multirow{1}{*}{Total bytes written to disk}         & $O(N \cdot E \cdot L \cdot T)$    & $O(N \cdot E \cdot L)$    & $O(N_\delta \cdot E \cdot L_\delta \cdot T)$  &   \Textcolor{myGreen}{\boldsymbol{$\blacktriangle$}}    & $O(N_\delta \cdot E \cdot L_\delta)$   &   \Textcolor{myGreen}{\boldsymbol{$\blacktriangle$}}    & $O(N \cdot E \cdot L \cdot T)$ & \Textcolor{myGrey}{\boldsymbol{$\bullet$}}   & $O(N \cdot E \cdot L)$ & \Textcolor{myGrey}{\boldsymbol{$\bullet$}}   & $O(N_\delta \cdot E \cdot L_\delta \cdot T)$  &   \Textcolor{myGreen}{\boldsymbol{$\blacktriangle$}}    & $O(N_\delta \cdot E \cdot L_\delta)$   &   \Textcolor{myGreen}{\boldsymbol{$\blacktriangle$}}   \\ 
        \midrule
        \multirow{1}{*}{Write amplification}         &  $O( L \cdot T)$   &  $O( L )$   &  $O( L \cdot T)$ &   \Textcolor{myGrey}{\boldsymbol{$\bullet$}}    & $O( L )$   &   \Textcolor{myGrey}{\boldsymbol{$\bullet$}}   & $O( L \cdot T)$ & \Textcolor{myGrey}{\boldsymbol{$\bullet$}}    & $O( L )$ & \Textcolor{myGrey}{\boldsymbol{$\bullet$}}  & $O( L \cdot T)$ &\Textcolor{myGrey}{\boldsymbol{$\bullet$}}  & $O( L )$ &\Textcolor{myGrey}{\boldsymbol{$\bullet$}} \\ 
        \midrule
        \multirow{1}{*}{Delete persistence latency}   & $O\left( \frac{T^{L-1} \cdot P \cdot B}{I} \right)$ & $O\left( \frac{T^L \cdot P \cdot B}{I} \right)$ & $O(\dpl)$ &   \Textcolor{myGreen}{\boldsymbol{$\blacktriangle$}}  & $O(\dpl)$  &   \Textcolor{myGreen}{\boldsymbol{$\blacktriangle$}}  & $O\left( \frac{T^{L-1} \cdot P \cdot B}{I} \right)$ & \Textcolor{myGrey}{\boldsymbol{$\bullet$}}   & $O\left( \frac{T^L \cdot P \cdot B}{I} \right)$ & \Textcolor{myGrey}{\boldsymbol{$\bullet$}}   & $O(\dpl)$ &   \Textcolor{myGreen}{\boldsymbol{$\blacktriangle$}}  & $O(\dpl)$  &   \Textcolor{myGreen}{\boldsymbol{$\blacktriangle$}}  \\
        \midrule
        \multirow{1}{*}{Zero result point lookup cost}   & $O( e^{-m/N})$ & $O( e^{-m/N} \cdot T)$ & $O( e^{-m/N_\delta} )$ &   \Textcolor{myGreen}{\boldsymbol{$\blacktriangle$}}  & $O( e^{-m/N_\delta} \cdot T )$ &   \Textcolor{myGreen}{\boldsymbol{$\blacktriangle$}}  & $O(h \cdot e^{-m/N})$   & \Textcolor{myRed}{\boldsymbol{$\blacktriangledown$}}   & $O( h \cdot e^{-m/N} \cdot T)$   & \Textcolor{myRed}{\boldsymbol{$\blacktriangledown$}} & $O(h \cdot e^{-m/N_\delta})$   & \Textcolor{myBlue}{\boldsymbol{$\blacklozenge$}}   & $O( h \cdot e^{-m/N_\delta} \cdot T)$   & \Textcolor{myBlue}{\boldsymbol{$\blacklozenge$}}  \\
        \midrule
        \multirow{1}{*}{Non-zero result point lookup cost}   & $O(1)$ & $O(1 + e^{-m/N} \cdot T)$ & $O(1)$ &   \Textcolor{myGrey}{\boldsymbol{$\bullet$}}  & $O(1 + e^{-m/N_\delta} \cdot T)$ &   \Textcolor{myGreen}{\boldsymbol{$\blacktriangle$}}  & $O(1 + h \cdot e^{-m/N})$   & \Textcolor{myRed}{\boldsymbol{$\blacktriangledown$}}   & $O( 1 + h \cdot e^{-m/N} \cdot T)$   &  \Textcolor{myRed}{\boldsymbol{$\blacktriangledown$}}  & $O(1 + h \cdot e^{-m/N_\delta})$   &  \Textcolor{myBlue}{\boldsymbol{$\blacklozenge$}}   & $O( 1 + h \cdot e^{-m/N_\delta} \cdot T)$   &  \Textcolor{myBlue}{\boldsymbol{$\blacklozenge$}}  \\
        \midrule
        \multirow{1}{*}{Short range point lookup cost}     & $O(L)$    & $O(L \cdot T)$    & $O(L_\delta)$  &   \Textcolor{myGreen}{\boldsymbol{$\blacktriangle$}}    & $O(L_\delta \cdot T)$ &   \Textcolor{myGreen}{\boldsymbol{$\blacktriangle$}}  & $O(h \cdot L)$  &   \Textcolor{myRed}{\boldsymbol{$\blacktriangledown$}}    & $O(h \cdot L \cdot T)$ &   \Textcolor{myRed}{\boldsymbol{$\blacktriangledown$}}  & $O(h \cdot L_\delta)$  &   \Textcolor{myBlue}{\boldsymbol{$\blacklozenge$}}    & $O(h \cdot L_\delta \cdot T)$ &   \Textcolor{myBlue}{\boldsymbol{$\blacklozenge$}}  \\ 
        \midrule
        \multirow{1}{*}{Long range point lookup cost}     & $O(\frac{s \cdot N}{B})$    & $O(\frac{T \cdot s \cdot N}{B})$    & $O(\frac{s \cdot N_{\delta}}{B})$  &   \Textcolor{myGreen}{\boldsymbol{$\blacktriangle$}}  & $O(\frac{T \cdot s \cdot N_{\delta}}{B})$ &   \Textcolor{myGreen}{\boldsymbol{$\blacktriangle$}}   & $O(\frac{s \cdot N}{B})$  &   \Textcolor{myGrey}{\boldsymbol{$\bullet$}}  & $O(\frac{T \cdot s \cdot N}{B})$ &   \Textcolor{myGrey}{\boldsymbol{$\bullet$}}    & $O(\frac{s \cdot N_{\delta}}{B})$  &   \Textcolor{myGreen}{\boldsymbol{$\blacktriangle$}}  & $O(\frac{T \cdot s \cdot N_{\delta}}{B})$ &   \Textcolor{myGreen}{\boldsymbol{$\blacktriangle$}}   \\ 
        \midrule
        \multirow{1}{*}{Insert/Update cost}          & $O(\tfrac{L \cdot T}{B})$    & $O(\tfrac{L}{B})$    & $O(\tfrac{L_\delta \cdot T}{B})$  &   \Textcolor{myGreen}{\boldsymbol{$\blacktriangle$}}   & $O(\tfrac{L_\delta}{B})$  &   \Textcolor{myGreen}{\boldsymbol{$\blacktriangle$}}   & $O(\tfrac{L \cdot T}{B})$  &   \Textcolor{myGrey}{\boldsymbol{$\bullet$}}   & $O(\tfrac{L}{B})$  &   \Textcolor{myGrey}{\boldsymbol{$\bullet$}}    & $O(\tfrac{L_\delta \cdot T}{B})$  &   \Textcolor{myGreen}{\boldsymbol{$\blacktriangle$}}   & $O(\tfrac{L_\delta}{B})$  &   \Textcolor{myGreen}{\boldsymbol{$\blacktriangle$}}   \\ 
        \midrule
        \multirow{1}{*}{Secondary range delete cost}     & $O(N/B)$    &  $O(N/B)$    &  $O(N_\delta/B)$   &   \Textcolor{myGreen}{\boldsymbol{$\blacktriangle$}}  &  $O(N_\delta/B)$    &   \Textcolor{myGreen}{\boldsymbol{$\blacktriangle$}} & $O\left( \tfrac{N}{B \cdot h} \right)$ & \Textcolor{myGreen}{\boldsymbol{$\blacktriangle$}}   & $O\left( \tfrac{N}{B \cdot h} \right)$  & \Textcolor{myGreen}{\boldsymbol{$\blacktriangle$}}    & $O\left( \tfrac{N_\delta}{B \cdot h} \right)$ & \Textcolor{myBlue}{\boldsymbol{$\blacklozenge$}}  & $O\left( \tfrac{N_\delta}{B \cdot h} \right)$ & \Textcolor{myBlue}{\boldsymbol{$\blacklozenge$}}   \\ 
        \midrule
        \multirow{1}{*}{Main memory footprint}     & $m + N \cdot \tfrac{k}{B}$    & $m + N \cdot \tfrac{k}{B}$        & $m + N_\delta \cdot \left( \tfrac{k}{B} + \tfrac{c}{B \cdot P} \right)$  & \Textcolor{myRed}{\boldsymbol{$\blacktriangledown$}}    & $m + N_\delta \cdot \left( \tfrac{k}{B} + \tfrac{c}{B \cdot P} \right)$  & \Textcolor{myRed}{\boldsymbol{$\blacktriangledown$}}    & $m + N \cdot \left( \tfrac{1}{B \cdot h} \right)$ & \Textcolor{myGreen}{\boldsymbol{$\blacktriangle$}}   & $m + N \cdot \left( \tfrac{1}{B \cdot h} \right)$ & \Textcolor{myGreen}{\boldsymbol{$\blacktriangle$}}  &   $m + N_\delta \cdot \left( \tfrac{k}{B \cdot h} + \tfrac{c}{B \cdot P} \right)$ & \Textcolor{myBlue}{\boldsymbol{$\blacklozenge$}}   &   $m + N_\delta \cdot \left( \tfrac{k}{B \cdot h} + \tfrac{c}{B \cdot P} \right)$  & \Textcolor{myBlue}{\boldsymbol{$\blacklozenge$}} \\ 
        \bottomrule
        \end{tabular}
    }
    \caption{Comparative analysis of state of the art and \comp{} (\hspace*{2mm}\Textcolor{myGreen}{\boldsymbol{$\blacktriangle$}}= better, \hspace*{2mm}\Textcolor{myRed}{\boldsymbol{$\blacktriangledown$}}= worse, \hspace*{2mm}\Textcolor{myGrey}{\boldsymbol{$\bullet$}}= same, \hspace*{2mm}\Textcolor{myBlue}{\boldsymbol{$\blacklozenge$}}= tunable). \label{tab:cost}}
    \vspace{-0.25in}
\end{table*}

\Paragraph{File Selection} 
\comp{} decides \textit{which} files to compact based on the trigger that invoked 
it. It has three modes: (i) the \textit{saturation-driven trigger 
and overlap-driven file selection} (SO), which is similar to the state of the art and
optimizes for write amplification, (ii) the \textit{saturation-driven trigger and 
delete-driven file selection} (SD), which selects the file with the highest $b$ to 
ensure
that as many tombstones as possible are compacted and to optimize for
space amplification, and (iii) the \textit{delete-driven trigger and delete-driven 
file selection} (DD), which selects a file with an expired tombstone to adhere to 
$\dpl$.
A tie in SD and DD is broken by picking the file that contains the oldest 
tombstone, and a tie in SO by picking the file with the most tombstones.
In case of a tie among levels, the smallest level is chosen 
for compaction to avoid write stalls during compaction. 
For a tie among files of the same level, \comp{} chooses 
the file with the most tombstones.

\vspace{-0.08in}
\subsubsection{\textbf{Implications on Performance}}
\comp{} guarantees that all point and range tombstones 
are persisted by the time their lifetime reaches $\dpl$ ($\forall f,\;a^{max}_f<\dpl$). We refer to the size of the tree as $N$ and 
to the size of the tree that has all entries persisted within $\dpl$ as $N_\delta$.

\Paragraph{Space amplification} \comp{} removes tombstones and 
logically invalidated entries from the tree on a rolling basis by compacting them in a 
time-bound fashion. By doing so, it diminishes the 
space amplification caused by out-of-place deletes, limiting $s_{amp}$ to $O(1/T)$ for 
leveling and $O(T)$ for tiering, even for a workload with deletes. 

\Paragraph{Write amplification} 
Ensuring delete persistence within $\dpl$, forces compactions 
on files with expired TTLs. Therefore, during a workload execution, initially 
\compshort{} leads to momentary spikes in write amplification. 
The initial high write amplification, however, is amortized over time. 
By eagerly compacting tombstones, 
\compshort{} purges most invalidated entries. Thus, 
future compactions involve fewer invalidated entries, leading to smaller write amplification 
which is comparable to the 
state of the art, as we show in Section \ref{sec:experimental_results}.

\Paragraph{Read performance} \compshort{} has a marginally positive effect on 
read performance. By compacting invalidated entries and point tombstones, \compshort{} 
reduces the number of entries hashed in the BFs, leading to smaller overall
FPR for a given amount of available memory, hence, 
the cost for point lookups on existing and non-existing keys is
improved asymptotically (Table \ref{tab:cost}). In the case that
$N_\delta$ entries can be stored in $L_\delta < L$ levels on disk, 
the lookup cost will benefit by accessing fewer levels.
Long range lookup cost is driven by the selectivity of the query, 
and this cost is lower for \compshort{} as timely persistence of deletes 
causes the query iterator to scan fewer invalidated entries.

\Paragraph{Persistence Guarantees} 
\compshort{} ensures that all tombstones inserted into an LSM-tree and 
flushed to the disk will always be compacted 
with the last level within the user-defined 
$\dpl$ threshold. 
Any tombstone retained in the write-ahead 
log (WAL) is consistently purged if the WAL is purged at a periodicity that 
is shorter than $\dpl$, which is typically the case in practice. 
Otherwise, we use a dedicated routine that checks all live 
WALs that are older than $\dpl$, copies all 
live records to a new WAL, and discards the records in the older WAL 
that made it to the disk.

\Paragraph{Practical Values for \boldsymbol{$\dpl$}} 
The delete persistence threshold of different applications vary widely. 
In commercial systems, LSM-engines are forced to 
perform a full tree compaction every 7, 30, or 60 days based on the 
SLA requirements~\cite{Huang2019}.

\Paragraph{Blind Deletes} 
A tombstone against a key that
does not exist or is already invalidated causes a \textit{blind delete}.
Blind deletes ingest tombstones against keys that 
do not exist in a tree, and these superfluous tombstones 
affect the performance of point and range queries~\cite{Huang2019}. 
To avoid blind deletes, \compshort{} probes the corresponding BF and inserts 
a tombstone only if the filter probe returns positive. This way, \compshort{} 
drastically reduces the number of blind deletes.

\vspace*{-3mm}
\subsection{\layout{} (\layoutshort{})} 
To facilitate secondary range deletes,
we introduce \textit{\layoutshort{}}, a continuum of physical storage layouts 
the arranges the data on disk in an \textbf{interweaved sorted order} on 
the sort key and delete key. \textit{\layoutshort{}} supports secondary 
range deletes without performing a full-tree compaction, at the cost of 
minimal extra metadata
and a tunable penalty on read performance.

\begin{figure*}[!ht]
    \centering  
    \hspace*{-5mm} 
        \includegraphics[scale=.23]{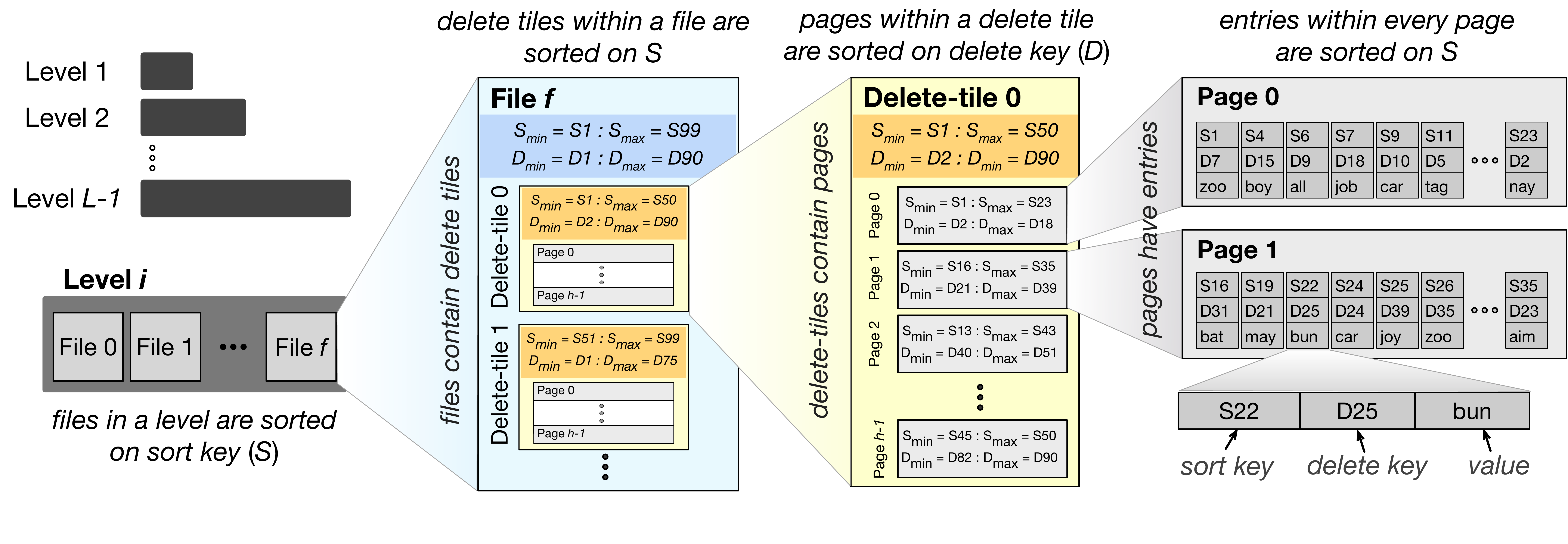} 
   \vspace*{-0.25in}        
    \caption{\layout{} stores data in an interweaved fashion on the sort and delete key to facilitate efficient secondary range deletes without hurting read performance.}
    \label{fig:layout} 
   \vspace*{-0.15in}        
\end{figure*}

\vspace{-0.1in}
\subsubsection{\textbf{The Layout}}
Figure \ref{fig:layout} presents the internal structure of \layoutshort{}. 
Essentially, \layoutshort{} adds one new layer in the storage layout of LSM trees.
In particular, in addition to the levels of the tree, the files of a level,
and the page of a file, we now introduce \textit{delete tiles} that belong
to a file and consist of pages. In the following discussion,
we use $\mathcal{S}$ to denote the sort key and $\mathcal{D}$ for the delete key. 

\Paragraph{Level Layout} 
The structure of the levels remains the same
as in state-of-the-art LSM trees. Every level is a collection of files 
containing non-overlapping ranges of $\mathcal{S}$. The order
between files in a level follows $\mathcal{S}$. Formally, if $i<j$, 
all entries in file $i$ have smaller $\mathcal{S}$ than those in file $j$.

\Paragraph{File Layout} 
The contents of the file are \textit{delete tiles},
which are collections of pages. Delete tiles contain non-overlapping ranges
of $\mathcal{S}$, hence from the perspective of the file, the order 
of the delete tiles also follows $\mathcal{S}$. Formally, if $k<l$, all entries in delete 
tile $k$ have smaller $\mathcal{S}$ than those in file $l$.

\Paragraph{Delete Tile Layout} 
Contrary to above, the pages of a delete tile
are sorted on $\mathcal{D}$. Formally, for $p<q$, page $p$ of a given delete
tile contains entries with smaller $\mathcal{D}$ than page $q$, while we 
have no information about $\mathcal{S}$. Organizing the contents of a tile
ordered on the delete key $\mathcal{D}$ allows us to quickly execute
range deletes because the entries under deletion are always clustered within
contiguous pages of each tile, which can be dropped in their entirety.

\Paragraph{Page layout} The order of entries \textit{within each page}
does not affect the performance of secondary range deletes, 
however, it significantly affects lookup cost, once a page is fetched to 
memory. To facilitate quick in-memory searches within a 
page~\cite{VanSandt2019}, we sort the entries of each page based on 
$\mathcal{S}$.

\vspace{-0.05in}
\subsubsection{\textbf{Facilitating Secondary Range Deletes}}  
\layoutshort{} exploits the fact that within a 
delete tile, the disk pages are sorted on the delete key. Hence, the entries 
targeted by a secondary range delete populate contiguous pages of each tile
(in the general case of every tile of the tree). The benefit of this approach
is that these pages can be dropped without having to be read and updated. 
Rather, they are removed from the otherwise immutable file and released to
be reclaimed by the underlying file system. We call this a \textit{full page
drop}. Pages containing entries at the edge of the delete range, might also
contain some valid entries. These pages
are loaded and the valid entries are identified with a tight \texttt{for}-loop 
on $\mathcal{D}$ (since they are sorted on $\mathcal{S}$). The cost of 
reading and re-writing these pages is the I/O cost of secondary
range deletes with \layoutshort{} when compared with a full-tree compaction
for the state of the art. We call these \textit{partial page drops}, and
constitute a small fraction of in-place editing, which is limited 
to zero or one page per delete tile. Subsequent compactions will
create files and delete tiles with the pre-selected sizes.

\vspace{-0.08in}
\subsubsection{\textbf{Tuning and Metadata}}
We now discuss the tuning knobs and the metadata of \layoutshort{}.

\Paragraph{Delete Tile Granularity} 
Every file contains a number of delete tiles, and each tile contains a 
number of pages. The basic tuning knob of \layoutshort{} is the number of 
pages per delete tile ($h$), which affects the granularity of delete ranges
that can be supported. 
For a file with $P$ disk pages, the number of delete tiles per file is $P/h$. 
The smallest granularity of a delete tile is when it consists of only a 
single disk page, i.e., for $h=1$. In fact, $h=1$ creates the same layout
as the state of the art, as all contents are sorted on $\mathcal{S}$ and
every range delete needs to update all data pages.
As $h$ increases, delete ranges with delete fraction close to $1/h$ will be
executed mostly by full drops. On the other hand, for higher
$h$ read performance is affected. The optimal decision for $h$ depends
on the workload (frequency of lookups and range deletes), and the
tuning (memory allocated to BFs and size ratio).

\Paragraph{Bloom Filters and Fence Pointers} 
We next discuss Bloom filters and fence pointers in the context of \layoutshort{}.

\textit{Bloom filters}. 
\layoutshort{} maintains BFs on $\mathcal{S}$ 
at \textit{the granularity of disk page}. 
Maintaining separate BFs per page requires no BF 
reconstruction for full page drops, and light-weight CPU cost for partial
page drops. The same overall FPR is achieved with the same memory 
consumption when having one BF per page, since a delete tile
contains no duplicates \cite{Athanassoulis2014}.

\textit{Fence pointers}. 
\layoutshort{} maintains fence pointers on $\mathcal{S}$
that keep track of the smallest sort key for every delete tile. 
Fence pointers on $\mathcal{S}$, aided by the BFs, accelerate 
lookups. To support secondary range deletes, for every delete tile, \layoutshort{}
maintains in memory a separate fence pointer structure on $\mathcal{D}$. 
We refer to this as 
\textit{delete fence pointers}. The delete fence pointers store the smallest 
$\mathcal{D}$ of every page enabling full page drops of the corresponding pages
without loading and searching the contents of a delete tile.

\Paragraph{Memory Overhead} 
While \layoutshort{} does not require any additional memory for BFs, it 
maintains two fence pointer structures -- one on $\mathcal{S}$ per delete tile and 
one on $\mathcal{D}$ per page. Since the state of the art maintains 
fence pointers on $\mathcal{S}$ per page, the space overhead for \layoutshort{} 
is the additional
metadata per tile. Assuming $sizeof(\mathcal{S})$ and $sizeof(\mathcal{D})$ 
are the sizes in bytes for $\mathcal{S}$ and $\mathcal{D}$ respectively,
the space overhead is: 
\small
\begin{eqnarray}
&\layoutshort{}_{mem}-SoA_{mem} = \nonumber \\
&\frac{N}{B\cdot h}\cdot sizeof(\mathcal{S}) + \frac{N}{B}\cdot sizeof(\mathcal{D}) - \frac{N}{B}\cdot sizeof(\mathcal{S})= \nonumber \\ 
& \#delete\_tiles\cdot\left(  sizeof\left(\mathcal{S}\right) + h\cdot \left( sizeof\left(\mathcal{D}\right) - sizeof\left(\mathcal{S}\right)\right) \right) \nonumber
\end{eqnarray}
\normalsize 
Note that if
$sizeof(\mathcal{S})=sizeof(\mathcal{D})$ the overhead is only one
sort key per tile, and if $sizeof(\mathcal{D})<sizeof(\mathcal{S})$,
\layoutshort{} might lead to less overall size of metadata.

\vspace{-0.05in}
\subsubsection{\textbf{CPU Overhead}}
\layoutshort{} navigates an intrinsic trade-off between the CPU cost for 
additional hashing for Bloom filters and the I/O cost associated with 
data movement to and from disk. For non-zero result point queries, \layoutshort{} performs 
$L\cdot h/4$ times more hash calculations compared to the state of the art, and 
$L\cdot h$ times in case of zero-result point queries. 
In practice, commercial LSM engines, such as RocksDB, use only 
a single MurmurHash hash digest to calculate which Bloom filter bits to 
set~\cite{RocksDB2018, Zhang2018a}. 
This reduces the overall cost of hash 
calculation by almost one order of magnitude.
We measured the time to hash a single 64-bit key using the MurmurHash to be 80$ns$, 
which is significantly smaller than the SSD access latency of 100$\mu$s. This allows 
\algo{} to strike a navigable trade-off between the CPU and I/O costs, and for the 
optimal value of $h$, \algo{} achieves a significantly superior overall performance 
as compared to the state of the art.

\vspace{-0.05in}
\subsubsection{\textbf{Implications on Performance}}
\layoutshort{} offers a tunable trade-off between the cost for 
secondary range deletes 
and that of lookups, but does not influence write performance (including space and 
write amplifications). 

\Paragraph{Point Lookup} 
A point read follows the same search algorithm as in the state of 
the art~\cite{Dayan2018a}.
In every level, a lookup searches the fence pointers (on 
$\mathcal{S}$) to locate the delete tile that may contain the search key. 
Once a delete tile is located, the BF for each delete tile page is 
probed. If a probe returns positive, the page is read to memory and 
binary searched, since the page is sorted on $\mathcal{S}$. If the key is found,
the query terminates. If not, the I/O was due to a false positive, 
and the next page of the tile is fetched. The I/O
cost for a query on an existing entry is $O(1 + h \cdot e^{-m/N})$ 
for leveling and $O(1 + T\cdot h \cdot e^{-m/N})$ for tiering. 
For a zero-result lookup the expected cost is $O(h \cdot e^{-m/N} )$   
and $O(h \cdot e^{-m/N} \cdot T)$ respectively.

\Paragraph{Range Lookup} In general, a range lookup may span several 
delete tiles spread in one or more consecutive files. 
\layoutshort{} affects the performance of range lookups only at the terminal delete 
tiles that contain the bounds on the range -- all delete tiles in between that are 
subsumed by the range always need to be read to memory. For the terminal delete tiles, 
the lookup needs to scan on average $h/2$ more pages per tile instead of only the 
qualifying pages. Thus, the cost for short range lookups for \layoutshort{} 
becomes $O(h \cdot L)$ for leveling and $O(h \cdot L \cdot T)$ for tiering. 
For long range lookups, the increase in cost gets amortized over the number 
of qualifying delete tiles, and remains the same asymptotically, 
i.e., $O(\tfrac{s \cdot N}{B})$ and $O(\tfrac{T \cdot s \cdot N}{B})$ for 
leveling and tiering, respectively. 

\Paragraph{Secondary Range Lookups} With the interweaved layout, 
\layoutshort{} can also support efficient range lookups on the delete key. While 
state-of-the-art LSM engines need to maintain a secondary index on the delete 
key, they still pay a heavy cost for scanning across many scattered pages. 
\layoutshort{} utilizes the ordering of the data on the delete key and can realize 
secondary range queries at a much lower I/O cost.

\subsubsection{\textbf{Navigable Design}}
A fundamental design goal for \layoutshort{} is to \textbf{navigate 
the trade-off} between the cost of \textit{secondary range deletes} and 
\textit{lookups}. 
\layoutshort{} offers a navigable continuum of storage layouts that can be 
tuned to obtain the optimal value for $h$ based on the workload 
characteristics and performance requirements. 
Assuming that the workload can be described
by the fractions of (a) point queries with zero-result {\small$f_{EPQ}$},
(b) point queries with non-zero result {\small$f_{PQ}$}, 
(c) short range queries {\small$f_{SRQ}$}, 
(d) long range queries {\small$f_{LRQ}$} with selectivity $s$, 
(e) secondary range deletes {\small$f_{SRD}$}, and 
(f) insertions {\small$f_{I}$}, then we can 
compare the cost of this workload for \algo{} and the state of the art.
\scriptsize
\begin{eqnarray}
	Cost_{\algo} \leq Cost_{SoA} \Rightarrow 
	f_{EPQ} \cdot FPR \cdot h +	f_{PQ} \cdot (1+\cdot FPR \cdot h) +\nonumber\\
	f_{SRQ} \cdot L_\delta\cdot h+ f_{LRQ} \cdot s \cdot {N}/{B} + f_{SRD}\cdot {N}/{(B\cdot h)}+ f_{I}\cdot log_T(N/B) \leq \nonumber\\ 	f_{EPQ} \cdot FPR +	f_{PQ} \cdot (1+\cdot FPR) +f_{SRQ} \cdot L\cdot + f_{LRQ} \cdot s \cdot \frac{N}{B} + \nonumber\\
	f_{SRD}\cdot {N}/{B}+ f_{I}\cdot log_T(N/B)	\Rightarrow \nonumber\\
	(f_{EPQ}+f_{PQ})\cdot FPR\cdot h+ f_{SRQ}\cdot L_\delta\cdot h + f_{SRD}\cdot N/(B\cdot h)\leq \nonumber\\ 	
		(f_{EPQ}+f_{PQ})\cdot FPR+ f_{SRQ}\cdot L + f_{SRD}\cdot N/B 
    \label{eqn:h}
    \end{eqnarray}
\normalsize
If we divide both sides of Eq. \ref{eqn:h} with $f_{SRD}$ we will get the costs with respect to
the relative frequencies of each operation with respect to the range deletes.
\scriptsize
\begin{eqnarray}
	(\ref{eqn:h}) \Rightarrow 
	(f_{EPQ}+f_{PQ})/f_{SRD}\cdot FPR\cdot (h-1)+ f_{SRQ}/f_{SRD}\cdot (L_\delta\cdot h -L)  \leq \nonumber\\  N/B \cdot (h - 1)/h \text{ (assuming }L_\delta\approx L\text{) }\nonumber\\
	\Rightarrow \frac{f_{EPQ}+f_{PQ}}{f_{SRD}}\cdot FPR+ \frac{f_{SRQ}}{f_{SRD}}\cdot L \leq \frac{N}{B} \cdot \frac{1}{h}
    \label{eqn:h2}
    \end{eqnarray}
\normalsize
Using Eq. \ref{eqn:h2} we navigate the secondary range delete vs. lookup performance trade-off
to find the best layout.

\subsection{\algo} 
\algo{} puts together the benefits of \comp{} and \layoutshort{} to better support deletes in 
LSM-trees. 
 For a given workload and a persistence delete latency threshold, \algo{} offers
 maximal performance, by carefully tuning the cost of persistent deletes, and
 their impact on the overall performance of the system. 
The key tuning knobs are (i) the 
delete persistence threshold ($\dpl$) and (ii) delete tile granularity ($h$). The 
delete persistence threshold is specified as part of the data retention SLA, and 
\algo{} sets the TTL for the tree-levels to ensure timely persistence. 

For a workload with secondary range deletes, \algo{} tunes the storage layout to find the optimal 
value for the delete tile granularity using the frequency of read operations relative to the
frequency of secondary range deletes. The optimal value of $h$ is given by solving Eq. \ref{eqn:h2}
with respect to $h$.

\vspace{-0.05in}
\small
\begin{equation}
    h \leq \frac{N}{B}\cdot \frac{1}{ \frac{f_{EPQ}+f_{PQ}}{f_{SRD}}\cdot FPR + \frac{f_{SRQ}}{f_{SRD}}\cdot L}  
   \label{eqn:h3}
\end{equation} 
\normalsize
\vspace{-0.05in}

For example, for a 400GB database and 4KB page size, if between two range deletes 
we execute 50M point queries of any type, 10K short range queries, 
and have $FPR\approx 0.02$ and $T=10$, using Eq. \ref{eqn:h3}, 
we have the optimal value of $h$ as:

\vspace{-0.1in}
\small
\begin{equation}
    h \leq \frac{{400GB/4KB}}{ 5\cdot 10^7\cdot 2\cdot 10^{-2} + 10^4 \cdot log_{10}(\frac{400GB}{4KB}) } = \frac{10^8}{ 10^6 + 8\cdot 10^4} \approx 10^2 .\nonumber
\end{equation} 
\normalsize

\Paragraph{Implementation} 
\algo{} is implemented on top of RocksDB which 
is an open-source LSM-based key-value store widely used in the
industry~\cite{Dong2017,FacebookRocksDB}. The current implementation of 
RocksDB is implemented as leveling (only Level 1 is implemented as tiered 
to avoid write-stalls) 
and has a fixed size ratio (defaults to $10$). 
 We develop a new API for \algo{} 
to have fine-grained control on the infrastructure. The API 
allows us to initiate compactions in RocksDB based on custom triggers 
and design custom file picking policies during compactions. 
RocksDB already stores metadata for every file, which includes the number of 
entries and deletes per files. We further store $age$ as the only additional
information per file. 
The delete persistence threshold is taken as a user-input at setup time and 
is used to dynamically set the level-TTLs.

%% file: 6-experimental_results.tex
 We evaluate \algo{} against state-of-the-art LSM-tree designs for a range of
 workloads with deletes and different delete persistence thresholds.

\Paragraph{Experimental Setup} We use a server with two 
Intel Xeon Gold 6230 2.1GHz processors each having 20 cores
with virtualization enabled and with 27.5MB L3 cache, 384GB of RDIMM main memory 
at 2933MHz and 240GB SSD.

\Paragraph{Default Setup}
Unless otherwise mentioned the experimental setup consists of
an initially empty database with ingestion rate at $2^{10}$ entries per second. 
The size of each entry is $1$KB, and the 
entries are uniformly and randomly distributed across the key domain and 
are inserted in random order. The size of the memory buffer is $1$MB 
(implemented as a skiplist). 
The size ratio for the database is set to $10$, and 
for the Bloom filters in memory, we use $10$ bits per entry.
To determine the 
raw performance, write operations are considered to 
have a lower priority than compactions. For all experiments performed, 
the implementation for \algo{} differed from the RocksDB setup in 
terms of only the compaction trigger and file picking policy. We have both block 
cache and direct I/O enabled and the WAL disabled. 
Deletes are issued only on keys that have been inserted in the database 
and are uniformly distributed within 
the workload. We insert 1GB data in the database with compactions 
given a higher priority than writes. 
The delete 
persistence threshold is set to $16.67\%$, 
$25\%$, and $50\%$ of the experiment's run-time. 
This experiment mimics the behavior of a long-running 
workload. The delete persistence threshold values
chosen for experimentation are representative of practical 
applications, where the threshold is $2$ months ($16.67\%$), 
$3$ months ($25\%$), $6$ months ($50\%$), respectively, 
for a commercial database running for $1$ 
year~\cite{Deshpande2018}. All 
lookups are issued after the whole database is populated.  

\Paragraph{Metrics}
The compaction related performance metrics including (i) \textit{total number of 
compactions performed}, (ii) \textit{total bytes compacted}, (iii) \textit{number 
of tombstones present in a tree}, and the (iv) \textit{age of files containing 
tombstones} are measured by taking a snapshot of the database after the 
experiment.
\textit{Space} and \textit{write amplification} are then computed using the 
equations from Sections \ref{subsubsec:SA} and \ref{subsubsec:WA}. 

\Paragraph{Workload} 
Given the lack of delete benchmarks, we designed a 
synthetic workload generator, which produces a variation of YCSB Workload A, 
with $50\%$ general updates and $50\%$ point lookups. In our experiments, we vary the 
percentage of deletes between $2\%$ to $10\%$ of the ingestion.

\subsection{Achieving Timely Delete Persistence}

\begin{figure*}[!ht]
    \centering
    \begin{subfigure}{0.25\textwidth}        
        \centering    
        \includegraphics[scale=0.33]{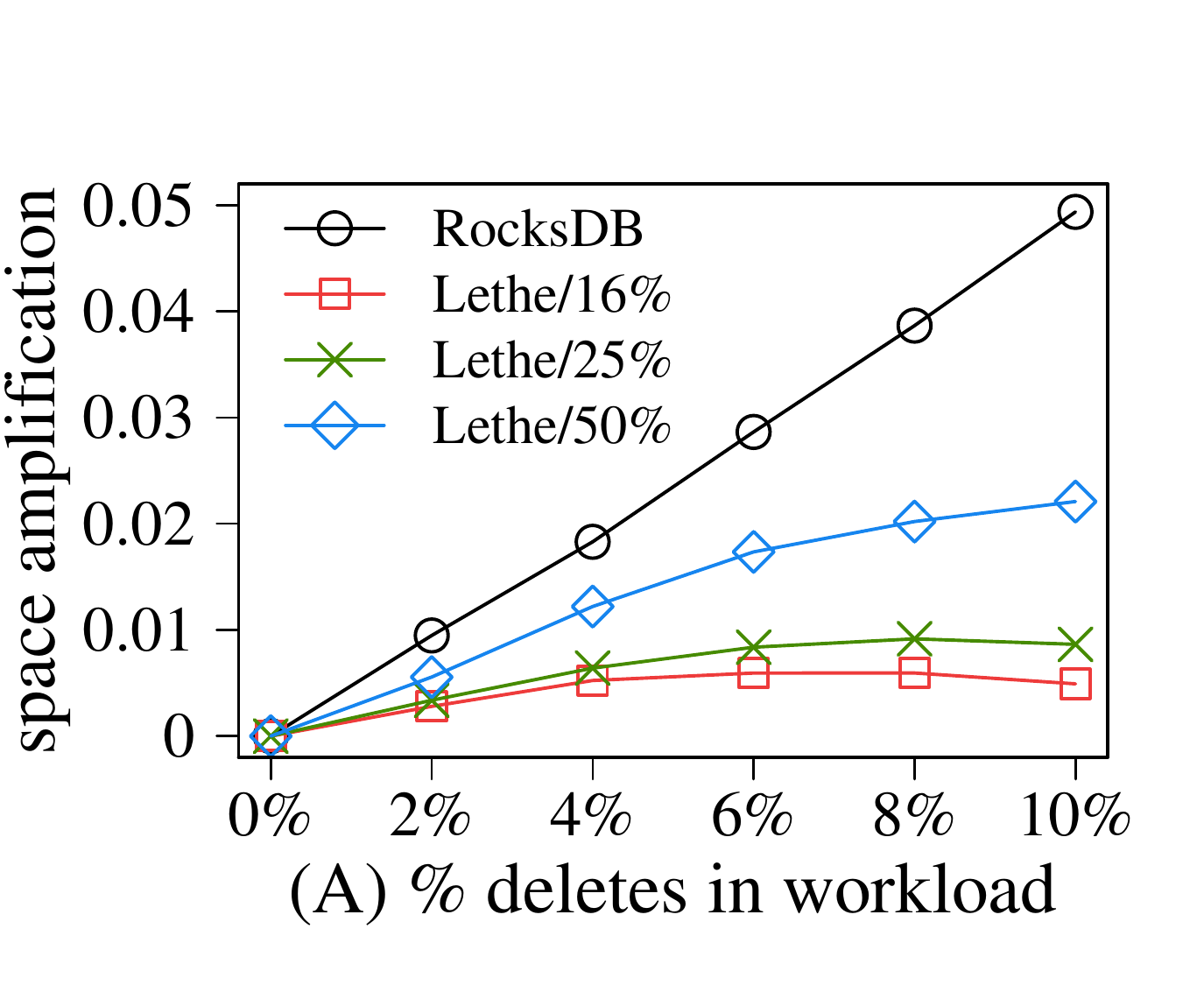}         
    \end{subfigure}%
    ~ \hspace*{-0.05in}
    \begin{subfigure}{0.25\textwidth} 
        \centering 
        \includegraphics[scale=0.33]{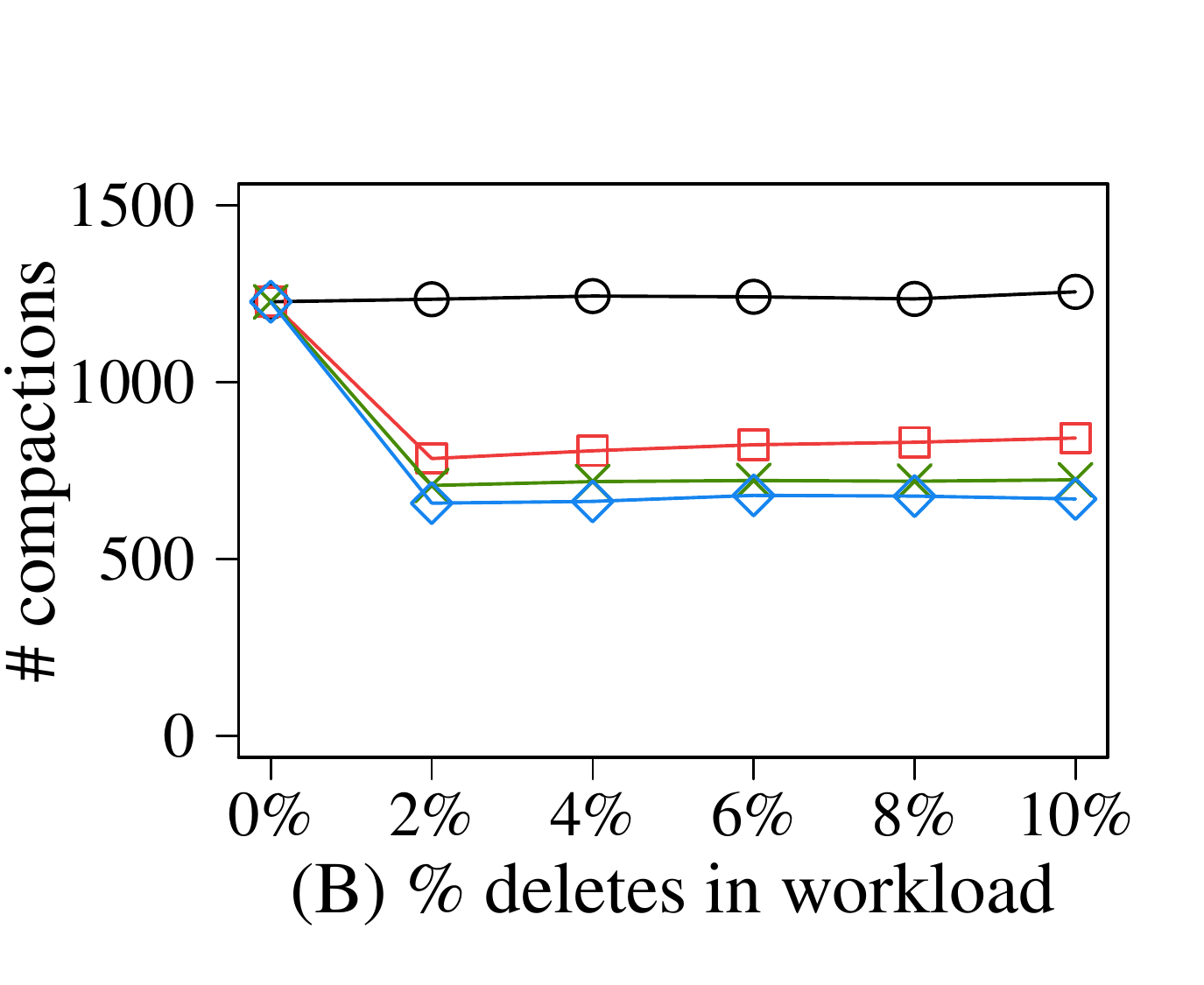}
    \end{subfigure}    
    ~ \hspace*{-0.05in}
    \begin{subfigure}{0.25\textwidth} 
        \centering
        \includegraphics[scale=0.33]{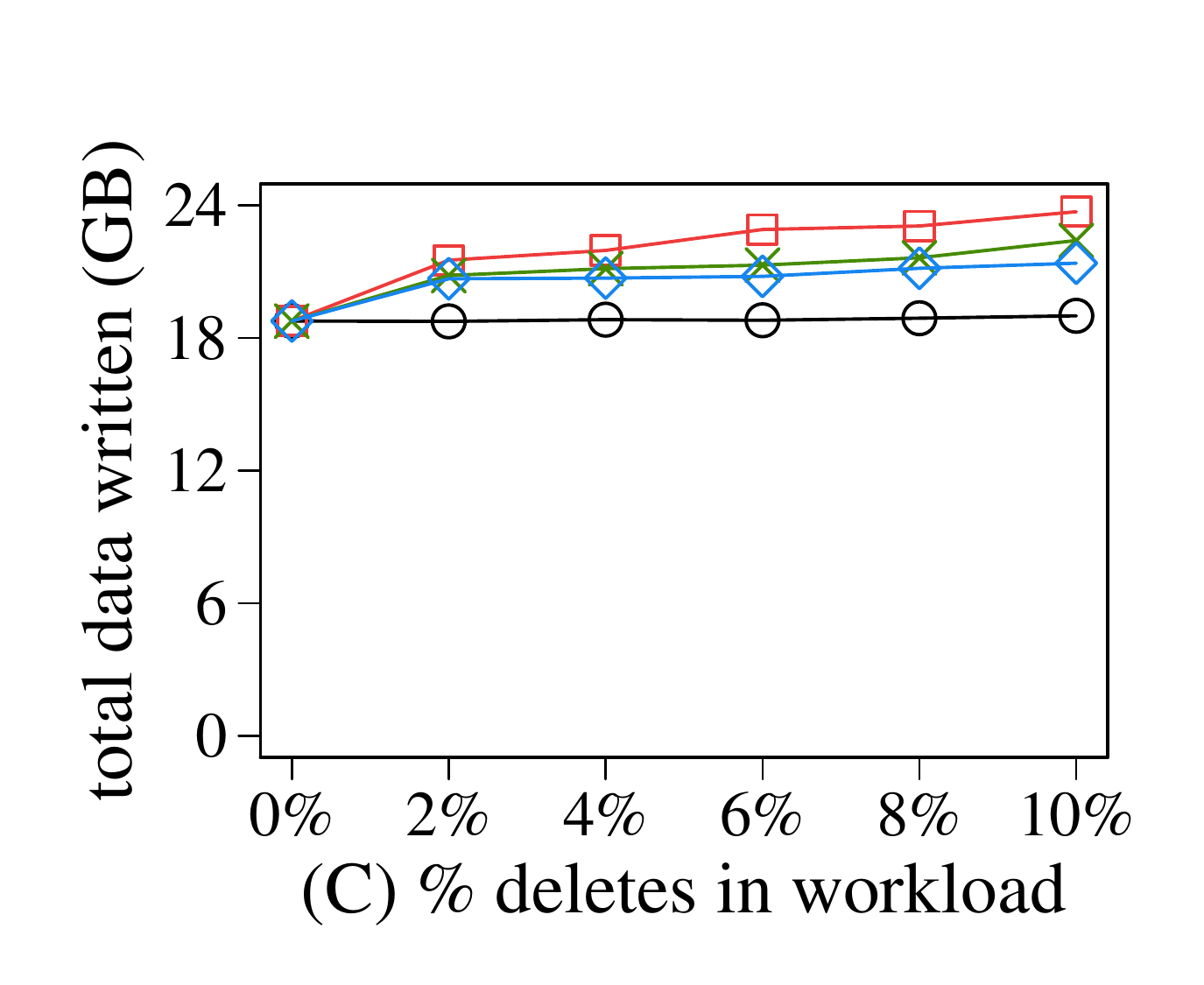}
    \end{subfigure}%
    ~ \hspace*{-0.05in}
    \begin{subfigure}{0.25\textwidth}        
        \centering    
        \includegraphics[scale=0.33]{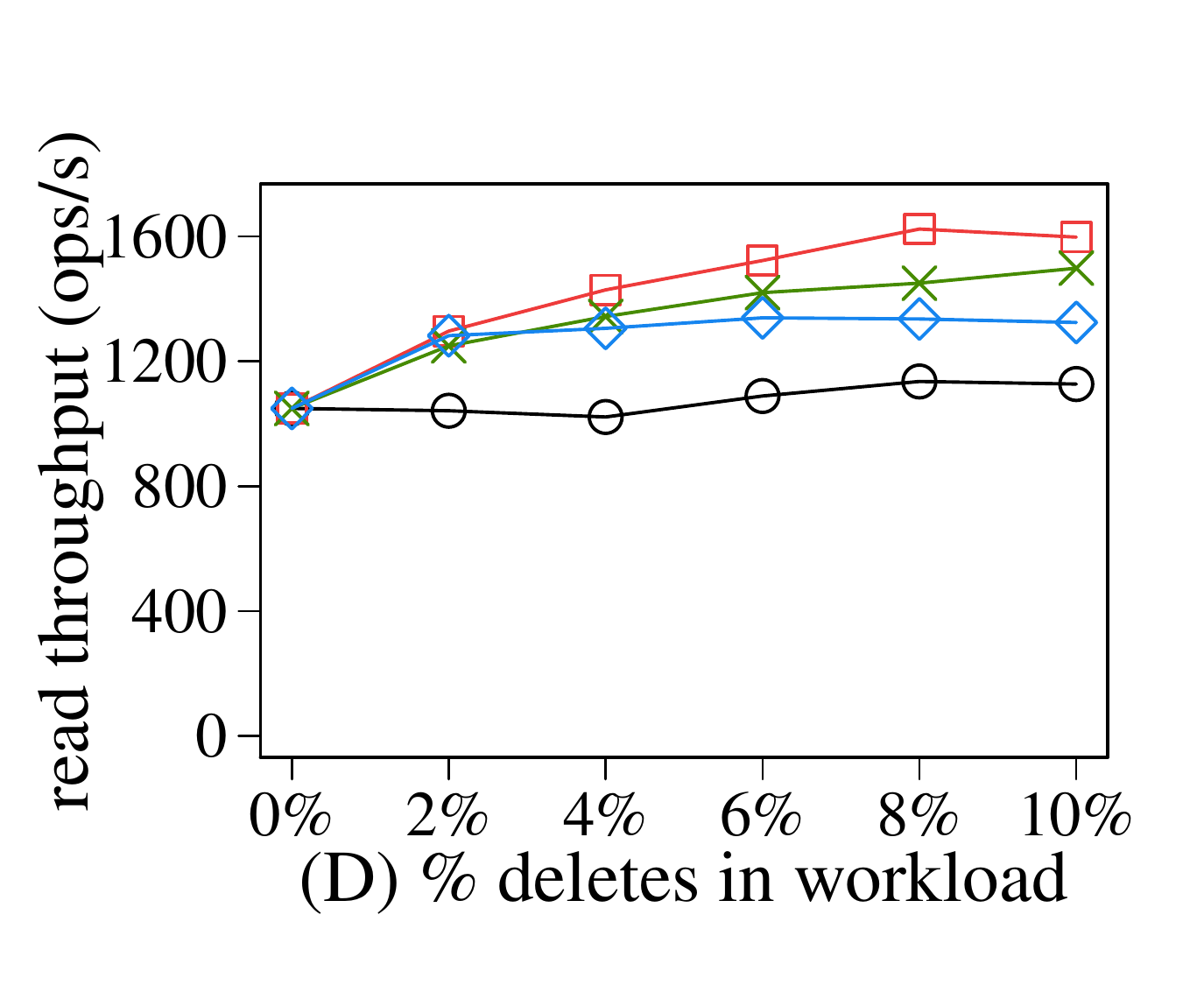}         
    \end{subfigure}%
\\
\vspace{-0.25in}
\hspace*{-3.5mm}
    \begin{subfigure}{0.25\textwidth}        
        \centering    \vspace*{2mm}
        \includegraphics[scale=0.33]{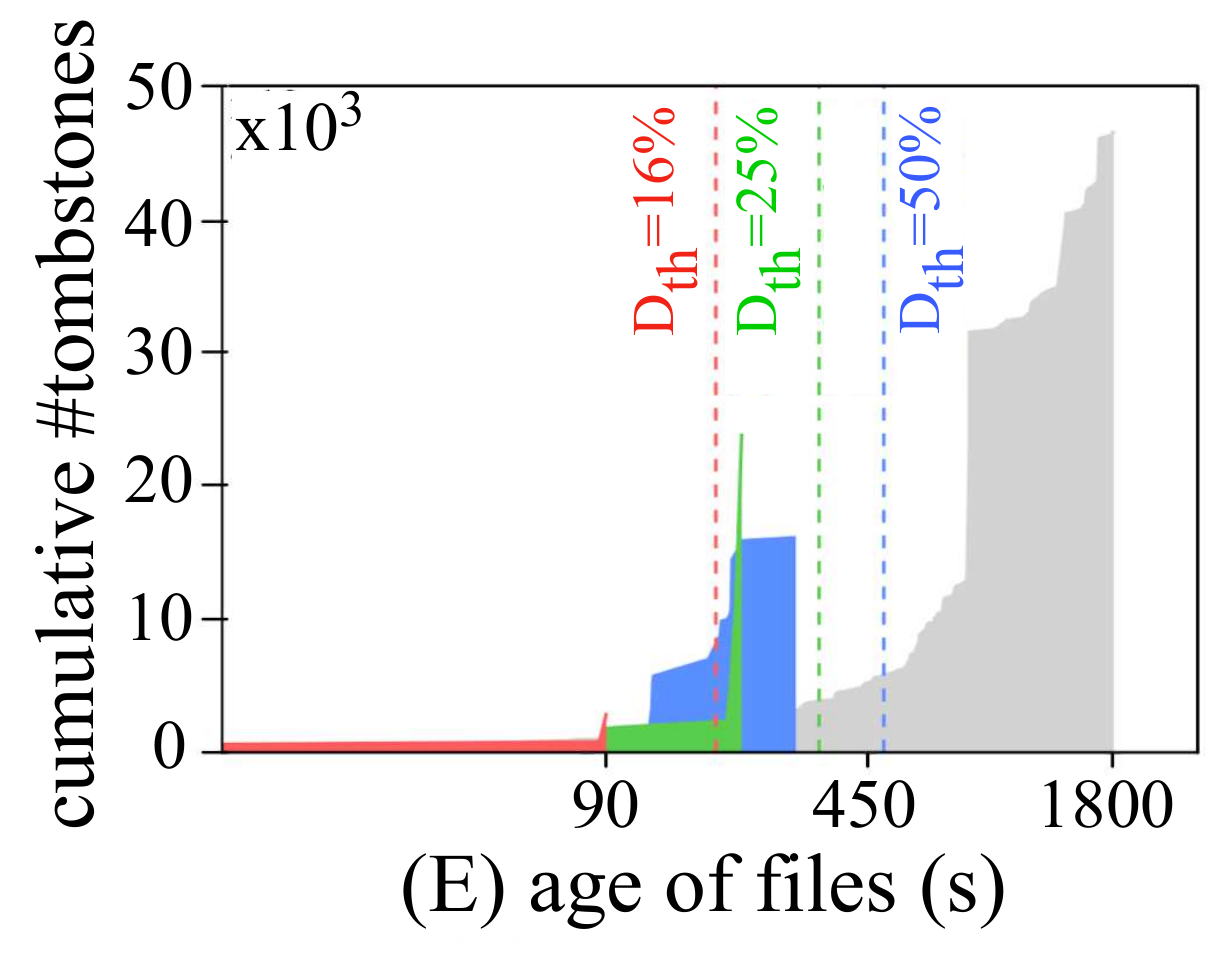}         
    \end{subfigure}%
    ~ \hspace*{-0.08in}  
    \begin{subfigure}{0.25\textwidth} 
        \centering 
        \includegraphics[scale=0.33]{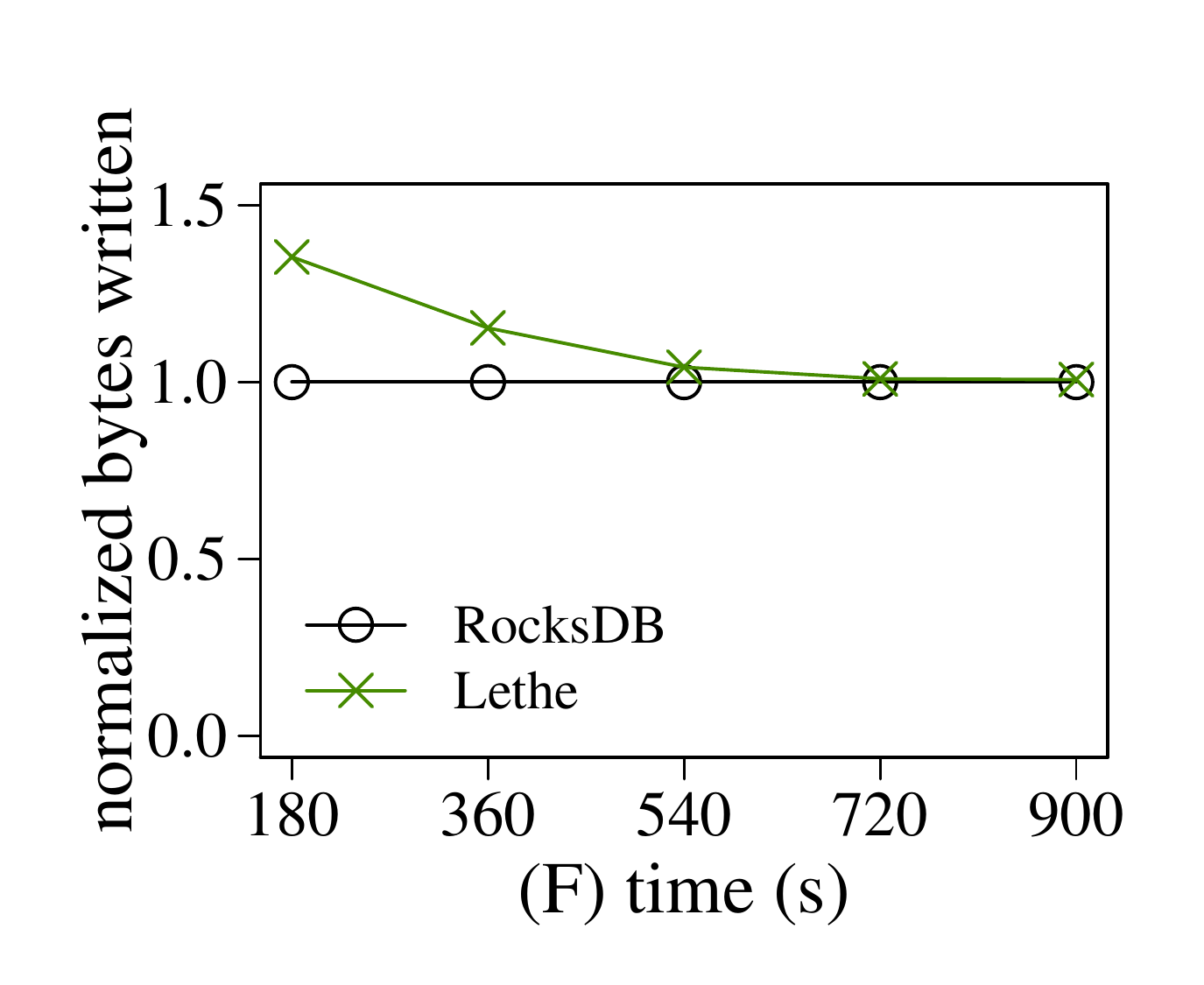}
    \end{subfigure}%
    ~ \hspace*{-0.02in}
    \begin{subfigure}{0.25\textwidth}        
        \centering    
        \includegraphics[scale=0.33]{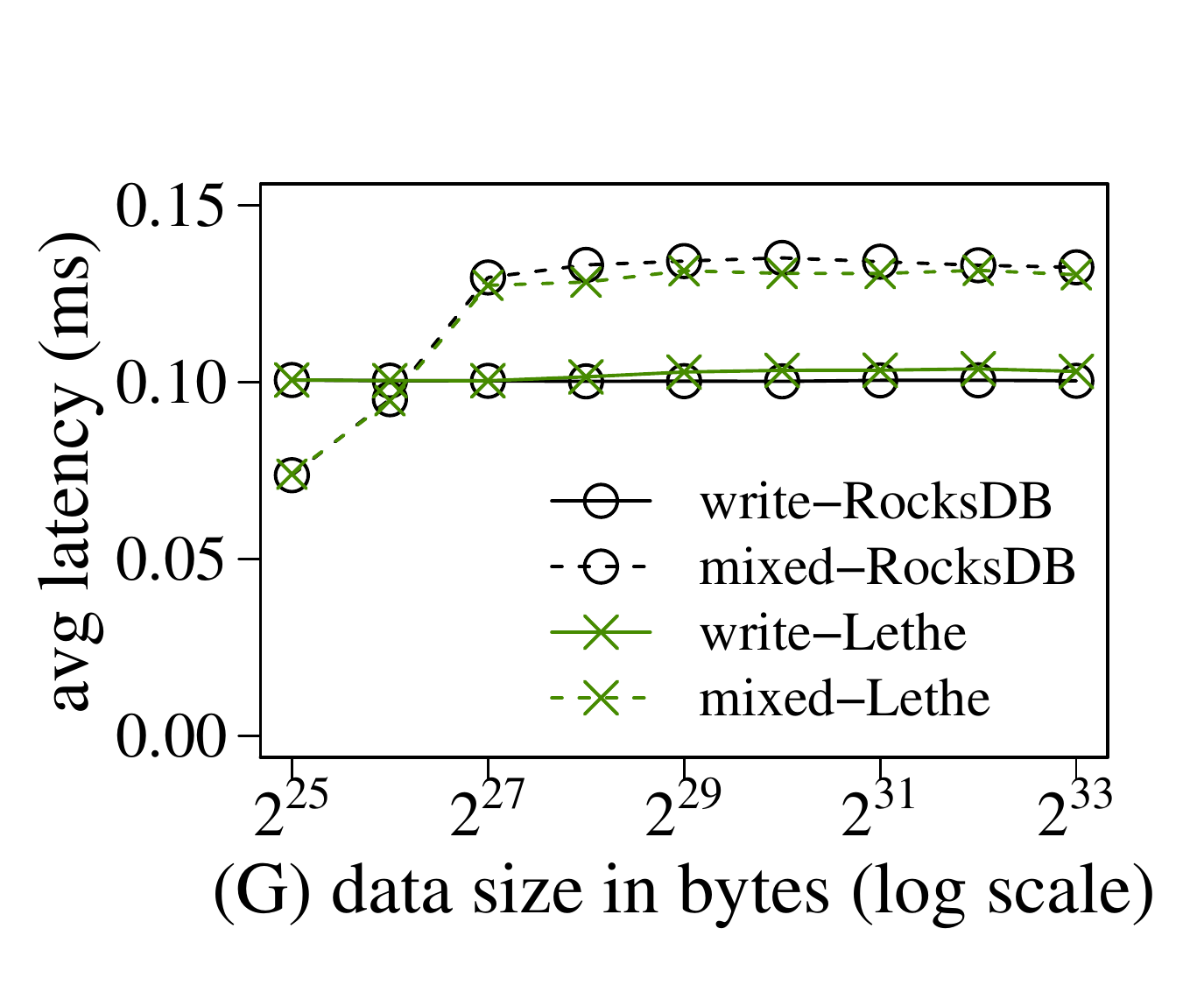}         
    \end{subfigure}%
    ~ \hspace*{-0.02in}
    \begin{subfigure}{0.25\textwidth}  
        \centering         
        \includegraphics[scale=0.33]{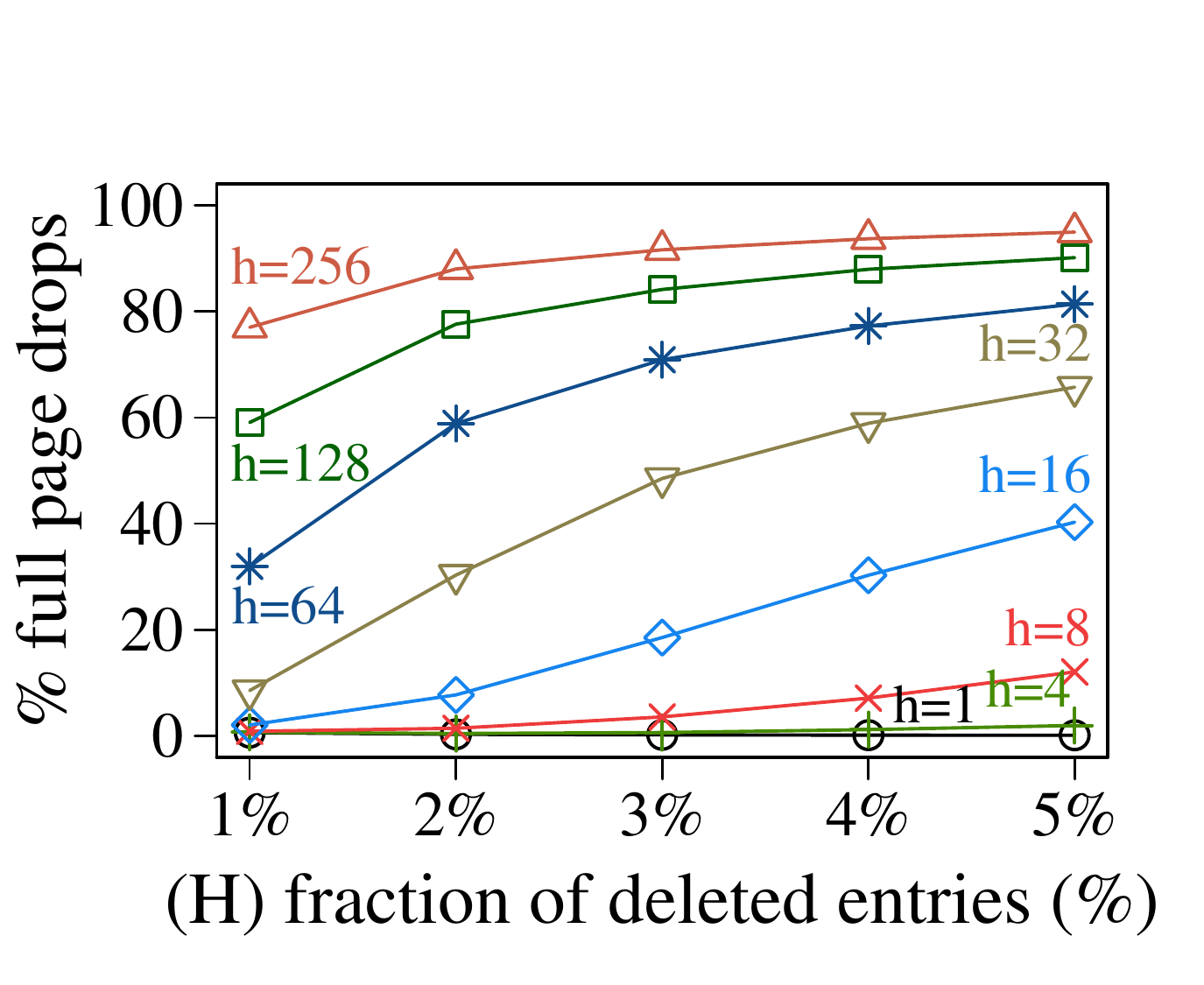}         
    \end{subfigure}%
    \\
    \vspace{-0.25in}
    \hspace*{-4.5mm}
    \begin{subfigure}{0.25\textwidth} 
        \centering   
        \includegraphics[scale=0.33]{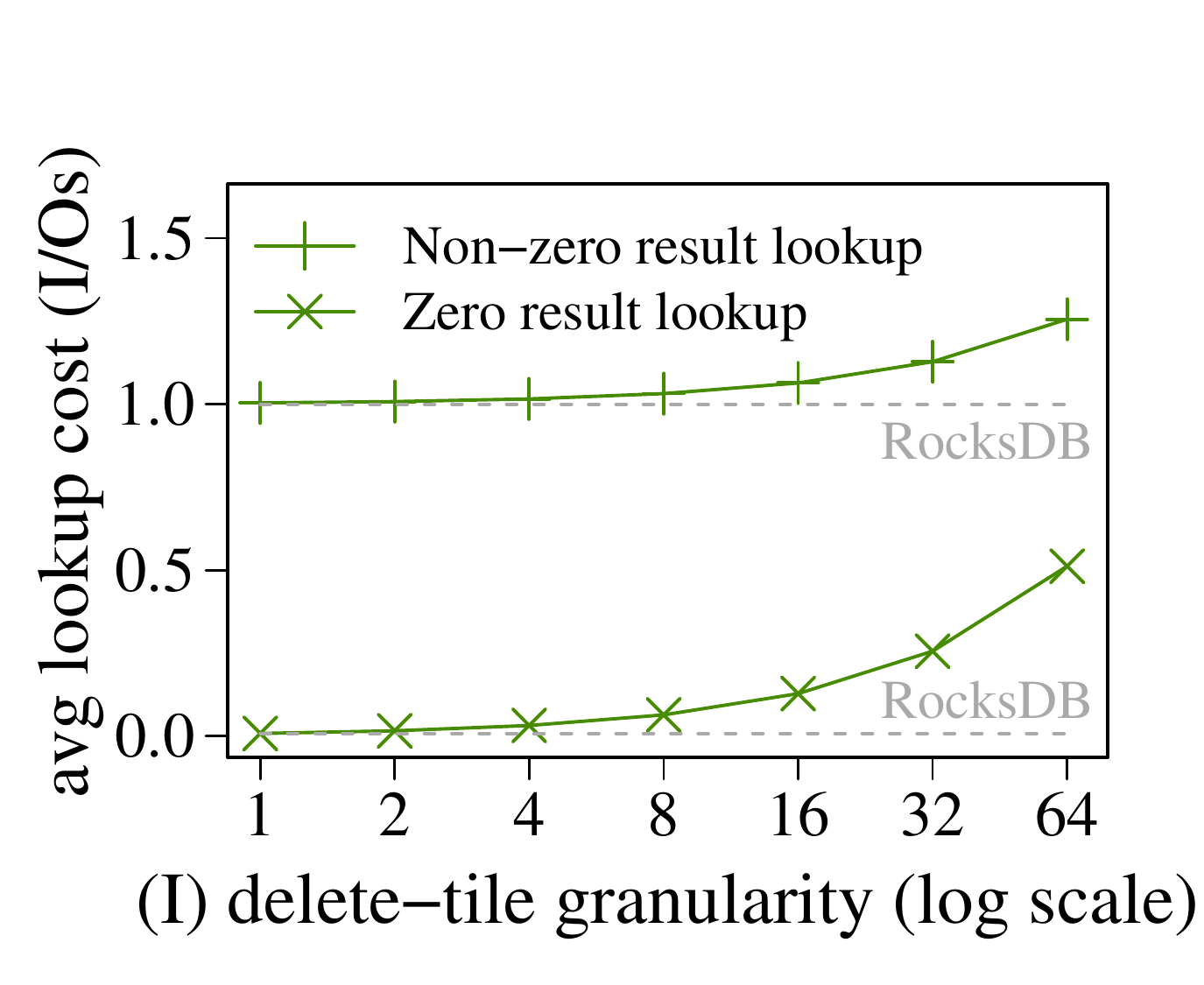}
    \end{subfigure}    
    ~ \hspace*{-0.08in}
    \begin{subfigure}{0.25\textwidth} 
        \centering
        \includegraphics[scale=0.33]{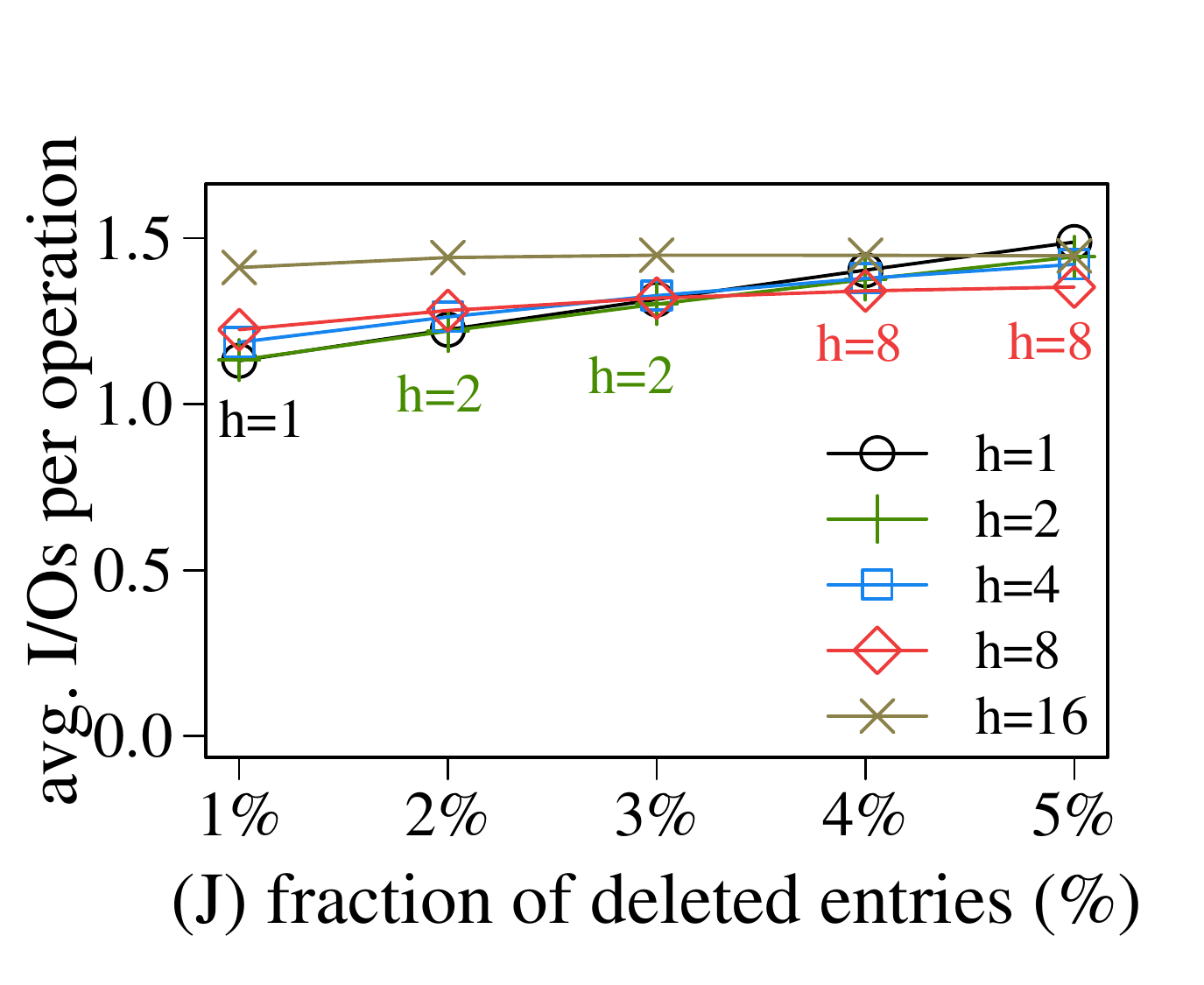}
    \end{subfigure}%
    ~ \hspace*{-0.1in}  
    \begin{subfigure}{0.25\textwidth}        
        \centering    
        \includegraphics[scale=0.33]{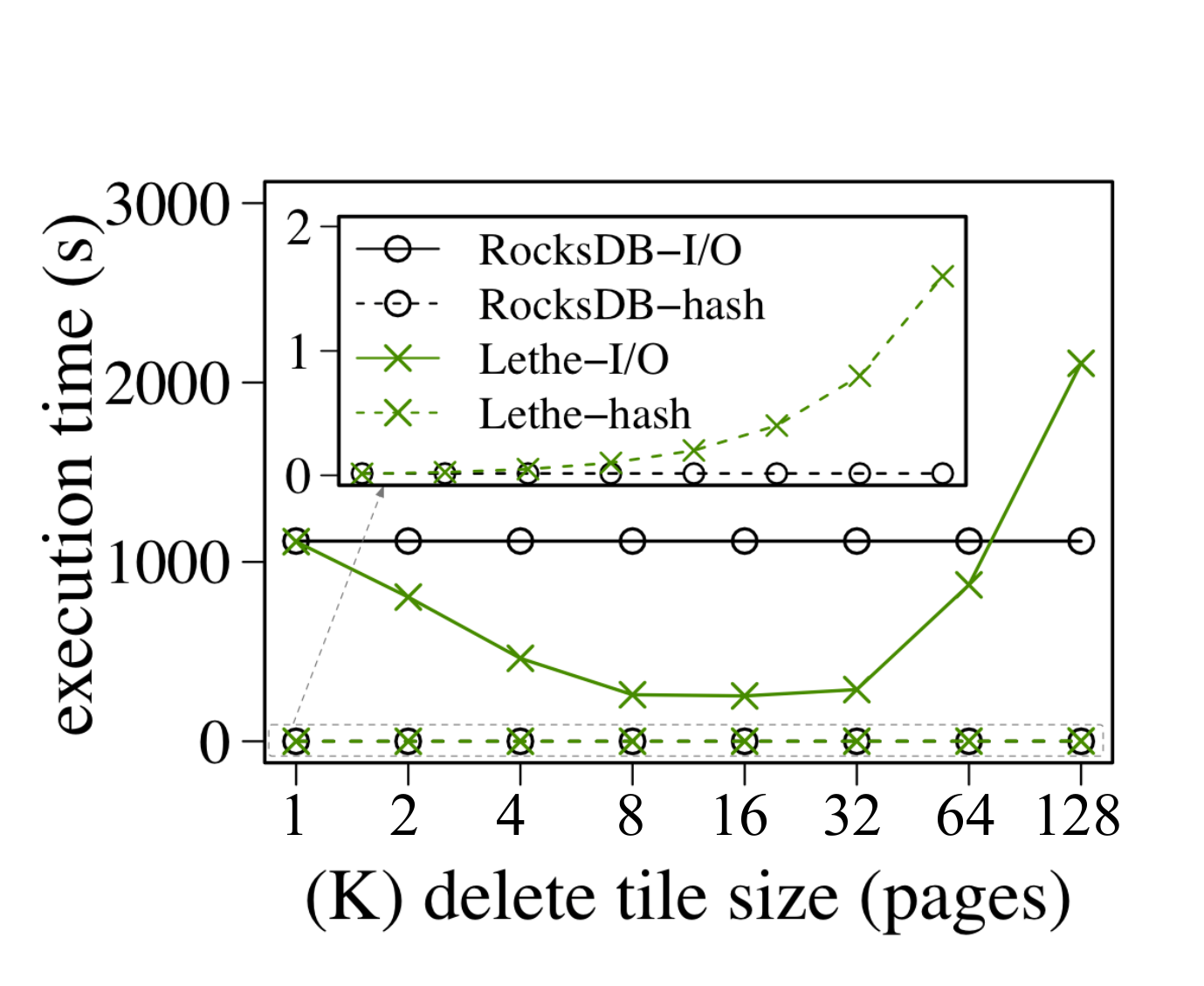}         
    \end{subfigure}%
    ~ \hspace*{-0.1in} 
    \begin{subfigure}{0.25\textwidth}        
        \centering    
        \includegraphics[scale=0.33]{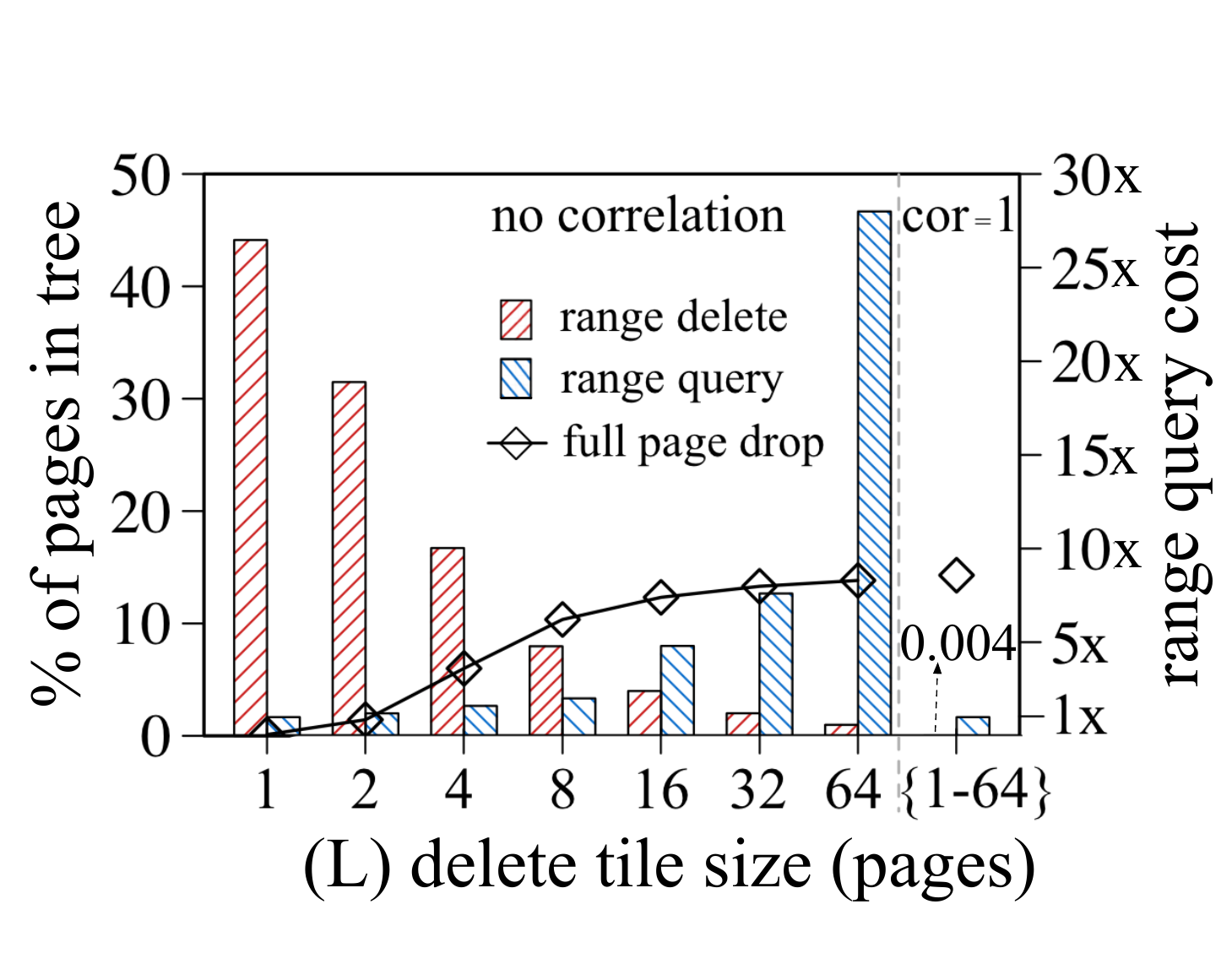}         
    \end{subfigure}%
    \vspace*{-3mm} 
    \caption{\algo{} improves space amplification (A) and performs 
    fewer larger compactions (B, C) to persist deletes timely (E). 
    In the process, it improves read throughput (D). The higher 
    write amplification in \algo{} gets amortized over time (F), which allows \algo{} 
    to scale similarly to RocksDB (G). (H) through (L) 
    show \algo{}'s ability to navigate the design space of storage layouts 
    and choose the optimal delete tile size to strike a balance between 
    the lookup and secondary range delete performance 
    to maximize the overall throughput. }  
    \label{fig:results} 
    \vspace*{-3mm}
\end{figure*}

\Paragraph{\algo{} Reduces Space Amplification}
We first show that \algo{} significantly reduces space amplification by 
persisting deletes timely. We set up this experiment by
varying the percentage of deletes in a workload for both RocksDB and \algo{}, 
for three different delete persistence thresholds. The plot is shown in
Figure \ref{fig:results} (A). For a workload with no deletes, the performances
of \algo{} and RocksDB are identical. This is because in the
\textit{absence of deletes}, \algo{} performs
compactions triggered by level-saturation, 
choosing files with minimal overlap.
In the \textit{presence of deletes}, driven by the delete
persistence threshold ($\dpl$), \algo{} compacts files more frequently to ensure
compliance with the threshold. It deletes the logically invalidated entries
persistently, and in the process, diminishes the space
amplification in the database. Even when $\dpl$ is set to $50\%$ of the
workload execution duration, \algo{} reduces space amplification
by about $48\%$. For shorter $\dpl$, the improvements in space amplification
are further pronounced by \algo{}.

\Paragraph{\algo{} Performs Fewer Compactions}
 Figures \ref{fig:results} (B) and (C) show that \algo{} performs fewer
 compactions as compared to RocksDB, but compacts more data during every
 compaction. \algo{} performs compactions on a rolling basis based on $D_{th}$.
 After each experiment, \algo{} was found to have fewer files on disk as
 compared to RocksDB. This is because, \algo{} compacts invalidated entries in a
 greedy manner, and for a workload with even a small fraction ($2\%$) of
 deletes, it reduces the number of compactions performed by $45\%$, as shown in
 Figure \ref{fig:results} (B). However, while compacting files with expired
 TTLs, the chosen file may overlap with a relatively higher number of files from
 the target level, and thus \algo{} compacts $4.5\%$ more data when $D_{th}$ is
 set as $50\%$ of the experiment's run-time, as shown in Figure \ref{fig:results} (C).

\Paragraph{\algo{} Achieves Better Read Throughput}
 In this experiment, we show that \algo{} offers a superior read performance as
 compared to RocksDB. For this experiment, we populate the database with $1$GB
 data and then issue point lookups on existing entries. Note that the lookups
 may be on entries have been deleted by a tombstone after they were inserted.
 With more deletes in the workload, the number of invalidated entries (including
 tombstones) hashed into the BFs increases. \algo{} purges these superfluous
 entries by persisting them in a time-bound manner, and thus, cleans up the BFs
 and improves their FPR. A lookup on a persistently deleted key returns negative
 without performing a disk I/O to read a tombstone. Overall, Figure
 \ref{fig:results} (D) shows that \algo{} improves lookup performance by up to
 $17\%$ for workloads with deletes.

\Paragraph{\algo{} Ensures Timely Delete Persistence} 
Figure \ref{fig:results} (E) shows the 
distribution of the tombstones ages at the end of the experiment to demonstrate
that \algo{} ensures timely persistent deletion. 
The X-axis shows the age of all files that contain tombstones, and the
Y-axis shows the cumulative number of tombstones at the instant of the 
snapshot with the age corresponding to the X-axis value or smaller. The goal
of \algo{} is to have fewer tombstones of smaller age than the state of the 
art, with all tombstones having age less than $D_{th}$
We show that in comparison with RocksDB, \algo{} persists 
between $40\%$ and $80\%$ more tombstones, and does so while honoring the delete persistence 
threshold. For $\dpl$= 50\% of the experiment's run-time, 
while RocksDB has $\sim 40,000$ tombstones (i.e., $\sim 40\%$ of all tombstones inserted) 
distributed among files that 
are older than $D_{th}$, \algo{} persists all deletes within the threshold.

\Paragraph{Write Amplification gets Amortized for \algo{}} 
This experiment demonstrates that the higher write amplification caused by 
the initial eager 
merging of \algo{} is amortized over time. For \algo{}, we set $\dpl$ to 60 seconds and 
take snapshots at an interval of 180 seconds during the execution of the experiments. 
At every snapshot, we measure the cumulative bytes written over the past intervals. 
We measure the same metric for RockDB (that does not support setting a $\dpl$), and use it to normalize the bytes written. 
We plot the normalized bytes written against time (across snapshots) 
in Figure \ref{fig:results} (F). 
We observe that due to eager merging, \algo{} writes $1.4\times$ more data 
compared to RocksDB in the first snapshot. However, by persisting invalid entries 
upfront, \algo{} purges superfluous entries from the tree, and hence, 
compacts fewer entries during subsequent compactions. This reduces the normalized 
writes by \algo{} over time. At the end 
of the experiment, we observe that \algo{} writes only $0.7\%$ more data as compared to 
RocksDB. In this experiment, we set the $\dpl$ to be $15\times$ smaller than 
the experiment duration to model the worst case. In practice, insertions in 
LSM engines continue for much longer (even perpetually) and $\dpl$ is 
set to a small constant duration. In this scenario \algo{}'s write amplification 
will be quickly amortized.

\Paragraph{\algo{} Scales Similarly to the State of the Art} This experiment 
shows that \algo{} and the state of the art follow the same performance trends
as data volume grows. 
We set up this experiment with the default configuration, and we vary 
data size. In addition to YCSB Workload A, which 
is used to compute the mixed workload latency, 
we use a write-only workload to measure write latency. 
Figure \ref{fig:results} (G), 
shows the average latency for both workloads with data size on the X-axis. We observe that 
the write latency for RocksDB and \algo{} is not affected by
data size. Due to the initial increased write amplification of \algo{}, 
its write latency is $0.1$-$3\%$ higher than that of RocksDB. For the mixed 
workload, however, \algo{} improves the average latency by $0.5$-$4\%$. This improvement 
is primarily due to the higher read throughput achieved by \algo{}, as shown in Figure 
\ref{fig:results} (D). For smaller data sizes, most data entries are stored in 
memory or the first disk level, which reduces read latency 
significantly.

 \vspace{-0.1in}
\subsection{Secondary Range Deletes}
Next, we evaluate secondary range deletes on \algo{}.

\Paragraph{Setup} 
Unless otherwise mentioned, the workload comprises $0.001\%$ secondary range delete 
operations along with $1\%$ range queries and $50\%$ point queries. Each file has $256$ pages and the 
size of every page is $4$KB. 

\Paragraph{\algo{} Achieves Superior Delete Performance}  
Figures \ref{fig:results} (H) through (L) show that 
\algo{} offers superior overall performance by storing the data on disk using \layoutshort. 
For the first experiment, we vary the selectivity of a 
secondary range delete operation, i.e., the fraction of the database 
that is deleted, and measure the number of pages that can be fully dropped 
during the operation. Full 
page drops do not require reading the page to memory, and thus, 
a higher value along the Y-axis is 
desirable. We repeat the experiment for different delete tile granularity ($h$). 
As the selectivity of the secondary range delete operation increases the 
number of full page drops decreases. This problem is further exacerbated 
for smaller delete tile granularity.

Although a higher value for $h$ is desirable for reducing the I/O requirement for secondary 
range deletes, it bears a trade-off with the lookup performance, as shown in 
Figure \ref{fig:results} (I). The cost of zero-result and non-zero result lookups increases linearly 
with an increase in the tile size. Thus, the optimal value for $h$ is driven by 
the workload composition. 

\Paragraph{Choosing the Optimal Storage Layout}
Figure \ref{fig:results} (J) shows \algo{}'s ability to navigate the continuum of storage layout and 
offer superior overall performance by determining the optimal storage layout. For a workload with a 
secondary range delete to point lookup ratio of $10^{-6}$ (i.e., $1$ secondary range delete per 
$0.1$M point lookups), as the selectivity of secondary range delete changes, the optimal way of 
storing data on disk changes. For selectivity $1\%$, the classical storage layout (with 
$h=1$) provides optimal performance. As the selectivity increases, the design choice changes, and 
we observe that for selectivity $5\%$ storing $8$ disk pages per delete tile ($h=8$) attains 
the optimal performance. 

\Paragraph{Analyzing the CPU-I/O Trade-off} 
In this experiment, we show the trade-off between the CPU and I/O costs 
for \algo{}. The workload for this experiment has 50\% point queries and 
1\% range query with selectivity of 0.001\%, 49\% inserts, and a single secondary range delete. We run this workload 
on a preloaded database of size $\sim$90GB (i.e., $2^{10}$ inserts/sec for 
24 hours). We have a single secondary range delete operation that deletes 1/7$^{th}$ 
of the database (mimics the behavior of deleting all data older than 7 days). 
We measure the total time spent for hash computation for filter probes and the 
total time spent for I/Os to the disk. 

Figure \ref{fig:results} (K) plots the total time spent in hashing and I/O access for both 
\algo{} and RocksDB while varying delete tile size. The embedded figure
shows that the hashing cost increases linearly with $h$. However, as the time 
to hash entries is 3 orders of magnitude smaller than the disk access latency and only point 
queries benefit from Bloom filters, the disk access time dominates 
the workload execution time. By design, \algo{} computes the 
optimal value of $h$, which in this case is 8. For $h=8$, the I/O cost for \algo{} 
is 76\% lower than that in RocksDB. This comes at the price of a $5\times$ 
increase in the hashing cost, which is completely hidden behind the massive
benefits in total number of I/Os.

\Paragraph{Effects of Correlation between Sort Key and Delete Key} 
Figure \ref{fig:results} (L) shows the effect of correlation between the sort key and the 
delete key. We run this experiment for two workloads. For the first workload, where 
there is no correlation between the sort and delete keys, the impact of the 
interweaved storage layout is prominent. As we increase $h$ the range delete cost 
drops drastically (because a larger fraction of pages can be fully dropped) at 
the expense of the cost of range queries. For the second workload, which has 
positive correlation between sort and delete key ($\approx$1), delete tiles have 
no impact on performance. For such a case, the classical LSM storage layout 
(i.e., $h=1$) becomes optimal.

%% file: 7-related_work.tex
\Paragraph{Deletion in Relational Systems}
Past work on data deletion on relational systems focuses on bulk
deletes~\cite{Bhattacharjee2007,Gartner2001,Lilja2007}. Efficient bulk
deletion relies on similar techniques as efficient reads: sorting or hashing data
to quickly locate, and ideally collocate, the entries to be deleted. Efficient
deletion has also been studied in the context of spatial data \cite{Lee1992} and view
maintenance \cite{Aravindan1997}. Contrary to past work, \algo{} aims to support 
a user-provided delete persistence latency threshold.

\Paragraph{Self-Destructing Data}
In addition, past research has proposed to automatically
make data disappear when specific conditions are met. Vanish is
a scheme that ensures that all copies of certain data become unreadable after a
user-specified time, without any specific action on the part of a
user~\cite{Geambasu2009, Geambasu2010}. Kersten \cite{Kersten2015} and
Heinis~\cite{Heinis2015} have proposed the concept of forgetting in data systems through
biology-inspired mechanisms as a way to better manage storage space and for efficient data analysis
capabilities, as the data generation trends continue to increase. Contrary to this, 
\algo{} supports timely data deletion that is set by users/applications.

%% file: 8-conclusion.tex
In this work, we show that state-of-the-art LSM-based key-value stores perform 
suboptimally for workloads with even a small proportion of deletes, and 
that the delete persistence latency in these data stores are potentially unbounded. 
To address this, we build \algo{}, a new LSM-based engine that introduces
a new family of compaction strategies \compshort{} and 
\layoutshort{}, a continuum of physical data storage layouts. \compshort{}
enforces delete persistence within a user-defined threshold while increasing
read throughput and reducing space amplification, at the expense of a modest increase in
 write amplification. \layoutshort{} offers efficient secondary range deletes, 
 which can be tuned
 to outperform state of the art for a given workload.
 
\vspace*{1mm}
\Paragraph{Acknowledgments} We thank the reviewers for their valuable feedback and 
researchers from Facebook for their useful remarks. We are particularly thankful to Xuntao Cheng from 
Alibaba and Mark Callaghan from MongoDB for the valuable discussions and concrete feedback. We also 
thank Zichen Zhu for helping with the experimentation. 
This work was partially funded by NSF Grant No. IIS-1850202.
\newpage

%% file: sigmod-archived.bbl
\begin{thebibliography}{10}

\bibitem{CCPA2018}
{California Consumer Privacy Act of 2018}.
\newblock {\em Assembly Bill No. 375, Chapter 55}, 2018.

\bibitem{Akidau2015}
T.~Akidau, R.~Bradshaw, C.~Chambers, S.~Chernyak, R.~Fern{\'{a}}ndez-Moctezuma,
  R.~Lax, S.~McVeety, D.~Mills, F.~Perry, E.~Schmidt, and S.~Whittle.
\newblock {The Dataflow Model: A Practical Approach to Balancing Correctness,
  Latency, and Cost in Massive-Scale, Unbounded, Out-of-Order Data Processing}.
\newblock {\em Proceedings of the VLDB Endowment}, 8(12):1792--1803, 2015.

\bibitem{Alagiannis2014}
I.~Alagiannis, S.~Idreos, and A.~Ailamaki.
\newblock {H2O: A Hands-free Adaptive Store}.
\newblock In {\em Proceedings of the ACM SIGMOD International Conference on
  Management of Data}, pages 1103--1114, 2014.

\bibitem{ApacheAccumulo}
Apache.
\newblock {Accumulo}.
\newblock {\em https://accumulo.apache.org/}.

\bibitem{ApacheCassandra}
Apache.
\newblock {Cassandra}.
\newblock {\em http://cassandra.apache.org}.

\bibitem{ApacheHBase}
Apache.
\newblock {HBase}.
\newblock {\em http://hbase.apache.org/}.

\bibitem{Appuswamy2017}
R.~Appuswamy, M.~Karpathiotakis, D.~Porobic, and A.~Ailamaki.
\newblock {The Case For Heterogeneous HTAP}.
\newblock In {\em Proceedings of the Biennial Conference on Innovative Data
  Systems Research (CIDR)}, 2017.

\bibitem{Aravindan1997}
C.~Aravindan and P.~Baumgartner.
\newblock {A Rational and Efficient Algorithm for View Deletion in Databases}.
\newblock In {\em Proceedings of the International Symposium on Logic
  Programming (ILPS)}, pages 165--179, 1997.

\bibitem{Arulraj2016}
J.~Arulraj, A.~Pavlo, and P.~Menon.
\newblock {Bridging the Archipelago between Row-Stores and Column-Stores for
  Hybrid Workloads}.
\newblock In {\em Proceedings of the ACM SIGMOD International Conference on
  Management of Data}, pages 583--598, 2016.

\bibitem{Athanassoulis2014}
M.~Athanassoulis and A.~Ailamaki.
\newblock {BF-Tree: Approximate Tree Indexing}.
\newblock {\em Proceedings of the VLDB Endowment}, 7(14):1881--1892, 2014.

\bibitem{Athanassoulis2011}
M.~Athanassoulis, S.~Chen, A.~Ailamaki, P.~B. Gibbons, and R.~Stoica.
\newblock {MaSM: Efficient Online Updates in Data Warehouses}.
\newblock In {\em Proceedings of the ACM SIGMOD International Conference on
  Management of Data}, pages 865--876, 2011.

\bibitem{Athanassoulis2015}
M.~Athanassoulis, S.~Chen, A.~Ailamaki, P.~B. Gibbons, and R.~Stoica.
\newblock {Online Updates on Data Warehouses via Judicious Use of Solid-State
  Storage}.
\newblock {\em ACM Transactions on Database Systems (TODS)}, 40(1), 2015.

\bibitem{Athanassoulis2016}
M.~Athanassoulis, M.~S. Kester, L.~M. Maas, R.~Stoica, S.~Idreos, A.~Ailamaki,
  and M.~Callaghan.
\newblock {Designing Access Methods: The RUM Conjecture}.
\newblock In {\em Proceedings of the International Conference on Extending
  Database Technology (EDBT)}, pages 461--466, 2016.

\bibitem{Bhattacharjee2007}
B.~Bhattacharjee, T.~Malkemus, S.~Lau, S.~Mckeough, J.-A. Kirton, R.~V.
  Boeschoten, and J.~Kennedy.
\newblock {Efficient Bulk Deletes for Multi Dimensionally Clustered Tables in
  DB2}.
\newblock In {\em Proceedings of the International Conference on Very Large
  Data Bases (VLDB)}, pages 1197--1206, 2007.

\bibitem{Callaghan2020}
M.~Callaghan.
\newblock {Deletes are fast and slow in an LSM}.
\newblock {\em
  http://smalldatum.blogspot.com/2020/01/deletes-are-fast-and-slow-in-lsm.html},
  2020.

\bibitem{Cao2020}
Z.~Cao, S.~Dong, S.~Vemuri, and D.~H.~C. Du.
\newblock {Characterizing, Modeling, and Benchmarking RocksDB Key-Value
  Workloads at Facebook}.
\newblock In {\em 18th USENIX Conference on File and Storage Technologies (FAST
  20)}, pages 209--223, 2020.

\bibitem{Chang2006}
F.~Chang, J.~Dean, S.~Ghemawat, W.~C. Hsieh, D.~A. Wallach, M.~Burrows,
  T.~Chandra, A.~Fikes, and R.~E. Gruber.
\newblock {Bigtable: A Distributed Storage System for Structured Data}.
\newblock In {\em Proceedings of the USENIX Symposium on Operating Systems
  Design and Implementation (OSDI)}, pages 205--218, 2006.

\bibitem{Cisco2018}
Cisco.
\newblock {Cisco Global Cloud Index: Forecast and Methodology, 2016–2021}.
\newblock {\em White Paper}, 2018.

\bibitem{Corbett2012}
J.~C. Corbett, J.~Dean, M.~Epstein, A.~Fikes, C.~Frost, J.~J. Furman,
  S.~Ghemawat, A.~Gubarev, C.~Heiser, P.~Hochschild, W.~C. Hsieh, S.~Kanthak,
  E.~Kogan, H.~Li, A.~Lloyd, S.~Melnik, D.~Mwaura, D.~Nagle, S.~Quinlan,
  R.~Rao, L.~Rolig, Y.~Saito, M.~Szymaniak, C.~Taylor, R.~Wang, and
  D.~Woodford.
\newblock {Spanner: Google's Globally-Distributed Database}.
\newblock In {\em Proceedings of the USENIX Symposium on Operating Systems
  Design and Implementation (OSDI)}, pages 251--264, 2012.

\bibitem{Dageville2016}
B.~Dageville, T.~Cruanes, M.~Zukowski, V.~Antonov, A.~Avanes, J.~Bock,
  J.~Claybaugh, D.~Engovatov, M.~Hentschel, J.~Huang, A.~W. Lee, A.~Motivala,
  A.~Q. Munir, S.~Pelley, P.~Povinec, G.~Rahn, S.~Triantafyllis, and
  P.~Unterbrunner.
\newblock {The Snowflake Elastic Data Warehouse}.
\newblock In {\em Proceedings of the ACM SIGMOD International Conference on
  Management of Data}, pages 215--226, 2016.

\bibitem{Dayan2017}
N.~Dayan, M.~Athanassoulis, and S.~Idreos.
\newblock {Monkey: Optimal Navigable Key-Value Store}.
\newblock In {\em Proceedings of the ACM SIGMOD International Conference on
  Management of Data}, pages 79--94, 2017.

\bibitem{Dayan2018a}
N.~Dayan, M.~Athanassoulis, and S.~Idreos.
\newblock {Optimal Bloom Filters and Adaptive Merging for LSM-Trees}.
\newblock {\em ACM Transactions on Database Systems (TODS)}, 43(4):16:1--16:48,
  2018.

\bibitem{Dayan2018}
N.~Dayan and S.~Idreos.
\newblock {Dostoevsky: Better Space-Time Trade-Offs for LSM-Tree Based
  Key-Value Stores via Adaptive Removal of Superfluous Merging}.
\newblock In {\em Proceedings of the ACM SIGMOD International Conference on
  Management of Data}, pages 505--520, 2018.

\bibitem{Dayan2019}
N.~Dayan and S.~Idreos.
\newblock {The Log-Structured Merge-Bush {\&} the Wacky Continuum}.
\newblock In {\em Proceedings of the ACM SIGMOD International Conference on
  Management of Data (SIGMOD)}, pages 449--466, 2019.

\bibitem{DeCandia2007}
G.~DeCandia, D.~Hastorun, M.~Jampani, G.~Kakulapati, A.~Lakshman, A.~Pilchin,
  S.~Sivasubramanian, P.~Vosshall, and W.~Vogels.
\newblock {Dynamo: Amazon's Highly Available Key-value Store}.
\newblock {\em ACM SIGOPS Operating Systems Review}, 41(6):205--220, 2007.

\bibitem{Deshpande2018}
A.~Deshpande and A.~Machanavajjhala.
\newblock {ACM SIGMOD Blog: Privacy Challenges in the Post-GDPR World: A Data
  Management Perspective}.
\newblock {\em http://wp.sigmod.org/?p=2554}, 2018.

\bibitem{Dong2017}
S.~Dong, M.~Callaghan, L.~Galanis, D.~Borthakur, T.~Savor, and M.~Strum.
\newblock {Optimizing Space Amplification in RocksDB}.
\newblock In {\em Proceedings of the Biennial Conference on Innovative Data
  Systems Research (CIDR)}, 2017.

\bibitem{FacebookMyRocks}
Facebook.
\newblock {MyRocks}.
\newblock {\em http://myrocks.io/}.

\bibitem{FacebookRocksDB}
Facebook.
\newblock {RocksDB}.
\newblock {\em https://github.com/facebook/rocksdb}.

\bibitem{Gartner2017}
Gartner.
\newblock {Gartner Says 8.4 Billion Connected ``Things" Will Be in Use in 2017,
  Up 31 Percent From 2016}.
\newblock https://tinyurl.com/Gartner2020, 2017.

\bibitem{Gartner2001}
A.~G{\"{a}}rtner, A.~Kemper, D.~Kossmann, and B.~Zeller.
\newblock {Efficient Bulk Deletes in Relational Databases}.
\newblock In {\em Proceedings of the IEEE International Conference on Data
  Engineering (ICDE)}, pages 183--192, 2001.

\bibitem{Geambasu2009}
R.~Geambasu, T.~Kohno, A.~A. Levy, and H.~M. Levy.
\newblock {Vanish: Increasing Data Privacy with Self-Destructing Data}.
\newblock In {\em Proceedings of the USENIX Security Symposium}, pages
  299--316, 2009.

\bibitem{Geambasu2010}
R.~Geambasu, A.~A. Levy, T.~Kohno, A.~Krishnamurthy, and H.~M. Levy.
\newblock {Comet: An active distributed key-value store}.
\newblock In {\em Proceedings of the USENIX Symposium on Operating Systems
  Design and Implementation (OSDI)}, pages 323--336, 2010.

\bibitem{Goddard2017}
M.~Goddard.
\newblock {The EU General Data Protection Regulation (GDPR): European
  Regulation that has a Global Impact}.
\newblock {\em International Journal of Market Research}, 59(6):703--705, 2017.

\bibitem{Golan-Gueta2015}
G.~Golan-Gueta, E.~Bortnikov, E.~Hillel, and I.~Keidar.
\newblock {Scaling Concurrent Log-Structured Data Stores}.
\newblock In {\em Proceedings of the ACM European Conference on Computer
  Systems (EuroSys)}, pages 32:1--32:14, 2015.

\bibitem{GoogleLevelDB}
Google.
\newblock {LevelDB}.
\newblock {\em https://github.com/google/leveldb/}.

\bibitem{Heinis2015}
T.~Heinis and A.~Ailamaki.
\newblock {Reconsolidating Data Structures}.
\newblock In {\em Proceedings of the International Conference on Extending
  Database Technology (EDBT)}, pages 665--670, 2015.

\bibitem{Heman2010}
S.~H{\'{e}}man, M.~Zukowski, and N.~J. Nes.
\newblock {Positional Update Handling in Column Stores}.
\newblock In {\em Proceedings of the ACM SIGMOD International Conference on
  Management of Data}, pages 543--554, 2010.

\bibitem{Huang2019}
G.~Huang, X.~Cheng, J.~Wang, Y.~Wang, D.~He, T.~Zhang, F.~Li, S.~Wang, W.~Cao,
  and Q.~Li.
\newblock {X-Engine: An Optimized Storage Engine for Large-scale E-commerce
  Transaction Processing}.
\newblock In {\em Proceedings of the ACM SIGMOD International Conference on
  Management of Data}, pages 651--665, 2019.

\bibitem{Huang2008}
Y.~Huang, T.~Z.~J. Fu, D.-M. Chiu, J.~C.~S. Lui, and C.~Huang.
\newblock {Challenges, design and analysis of a large-scale p2p-vod system}.
\newblock In {\em Proceedings of the ACM SIGCOMM 2008 Conference on
  Applications, Technologies, Architectures, and Protocols for Computer
  Communications, Seattle, WA, USA, August 17-22, 2008}, pages 375--388, 2008.

\bibitem{Hueske2018}
F.~Hueske.
\newblock {State TTL for Apache Flink: How to Limit the Lifetime of State}.
\newblock {\em Ververica}, 2018.

\bibitem{Kersten2015}
M.~L. Kersten.
\newblock {Big Data Space Fungus}.
\newblock In {\em Proceedings of the Biennial Conference on Innovative Data
  Systems Research (CIDR), Gong show talk}, 2015.

\bibitem{Kulkarni2015}
S.~Kulkarni, N.~Bhagat, M.~Fu, V.~Kedigehalli, C.~Kellogg, S.~Mittal, J.~M.
  Patel, K.~Ramasamy, and S.~Taneja.
\newblock {Twitter Heron: Stream Processing at Scale}.
\newblock In {\em Proceedings of the ACM SIGMOD International Conference on
  Management of Data}, pages 239--250, 2015.

\bibitem{Lamb2012}
A.~Lamb, M.~Fuller, and R.~Varadarajan.
\newblock {The Vertica Analytic Database: C-Store 7 Years Later}.
\newblock {\em Proceedings of the VLDB Endowment}, 5(12):1790--1801, 2012.

\bibitem{Lee1992}
J.-T. Lee and G.~G. Belford.
\newblock {An Efficient Object-based Algorithm for Spatial Searching, Insertion
  and Deletion}.
\newblock In {\em Proceedings of the IEEE International Conference on Data
  Engineering (ICDE)}, pages 40--47, 1992.

\bibitem{Li2010}
Y.~Li, B.~He, J.~Yang, Q.~Luo, K.~Yi, and R.~J. Yang.
\newblock {Tree Indexing on Solid State Drives}.
\newblock {\em Proceedings of the VLDB Endowment}, 3(1-2):1195--1206, 2010.

\bibitem{Lilja2007}
T.~Lilja, R.~Saikkonen, S.~Sippu, and E.~Soisalon-Soininen.
\newblock {Online Bulk Deletion}.
\newblock In {\em Proceedings of the IEEE International Conference on Data
  Engineering (ICDE)}, pages 956--965, 2007.

\bibitem{LinkedInVoldemort}
LinkedIn.
\newblock {Voldemort}.
\newblock {\em http://www.project-voldemort.com}.

\bibitem{Lu2004}
H.~Lu.
\newblock {Peer-to-Peer Support for Massively Multiplayer Games}.
\newblock In {\em Proceedings IEEE INFOCOM 2004, The 23rd Annual Joint
  Conference of the IEEE Computer and Communications Societies, Hong Kong,
  China, March 7-11, 2004}, 2004.

\bibitem{Luo2019}
C.~Luo and M.~J. Carey.
\newblock {LSM-based Storage Techniques: A Survey}.
\newblock {\em The VLDB Journal}, 2019.

\bibitem{Mohan2016}
C.~Mohan.
\newblock {Hybrid Transaction and Analytics Processing (HTAP): State of the
  Art}.
\newblock In {\em Proceedings of the International Workshop on Business
  Intelligence for the Real-Time Enterprise (BIRTE)}, 2016.

\bibitem{MongoDB}
MongoDB.
\newblock {Online reference}.
\newblock {\em http://www.mongodb.com/}.

\bibitem{ONeil1996}
P.~E. O'Neil, E.~Cheng, D.~Gawlick, and E.~J. O'Neil.
\newblock {The log-structured merge-tree (LSM-tree)}.
\newblock {\em Acta Informatica}, 33(4):351--385, 1996.

\bibitem{Ozcan2017}
F.~{\"{O}}zcan, Y.~Tian, and P.~T{\"{o}}z{\"{u}}n.
\newblock {Hybrid Transactional/Analytical Processing: A Survey}.
\newblock In {\em Proceedings of the ACM SIGMOD International Conference on
  Management of Data}, pages 1771--1775, 2017.

\bibitem{Pantelopoulos2010}
A.~Pantelopoulos and N.~G. Bourbakis.
\newblock {Prognosis: a wearable health-monitoring system for people at risk:
  methodology and modeling}.
\newblock {\em IEEE Trans. Information Technology in Biomedicine},
  14(3):613--621, 2010.

\bibitem{Papadopoulos2016}
S.~Papadopoulos, K.~Datta, S.~Madden, and T.~Mattson.
\newblock {The TileDB Array Data Storage Manager}.
\newblock {\em Proceedings of the VLDB Endowment}, 10(4):349--360, 2016.

\bibitem{Raju2017}
P.~Raju, R.~Kadekodi, V.~Chidambaram, and I.~Abraham.
\newblock {PebblesDB: Building Key-Value Stores using Fragmented Log-Structured
  Merge Trees}.
\newblock In {\em Proceedings of the ACM Symposium on Operating Systems
  Principles (SOSP)}, pages 497--514, 2017.

\bibitem{RocksDB2018}
RocksDB.
\newblock {DeleteRange: A New Native RocksDB Operation}.
\newblock {\em https://rocksdb.org/blog/2018/11/21/delete-range.html}, 2018.

\bibitem{Sakaki2010}
T.~Sakaki, M.~Okazaki, and Y.~Matsuo.
\newblock {Earthquake shakes Twitter users: real-time event detection by social
  sensors}.
\newblock In {\em Proceedings of the 19th International Conference on World
  Wide Web, WWW 2010, Raleigh, North Carolina, USA, April 26-30, 2010}, pages
  851--860, 2010.

\bibitem{Sarkar2018}
S.~Sarkar, J.-P. Ban{\^{a}}tre, L.~Rilling, and C.~Morin.
\newblock {Towards Enforcement of the EU GDPR: Enabling Data Erasure}.
\newblock In {\em Proceedings of the IEEE International Conference of Internet
  of Things (iThings)}, pages 1--8, 2018.

\bibitem{Sears2012}
R.~Sears and R.~Ramakrishnan.
\newblock {bLSM: A General Purpose Log Structured Merge Tree}.
\newblock In {\em Proceedings of the ACM SIGMOD International Conference on
  Management of Data}, pages 217--228, 2012.

\bibitem{SQLite4}
SQLite4.
\newblock {Online reference}.
\newblock {\em https://sqlite.org/src4/}.

\bibitem{Stonebraker2005}
M.~Stonebraker, D.~J. Abadi, A.~Batkin, X.~Chen, M.~Cherniack, M.~Ferreira,
  E.~Lau, A.~Lin, S.~R. Madden, E.~J. O'Neil, P.~E. O'Neil, A.~Rasin, N.~Tran,
  and S.~Zdonik.
\newblock {C-Store: A Column-oriented DBMS}.
\newblock In {\em Proceedings of the International Conference on Very Large
  Data Bases (VLDB)}, pages 553--564, 2005.

\bibitem{Thonangi2017}
R.~Thonangi and J.~Yang.
\newblock {On Log-Structured Merge for Solid-State Drives}.
\newblock In {\em Proceedings of the IEEE International Conference on Data
  Engineering (ICDE)}, pages 683--694, 2017.

\bibitem{To2017}
Q.-C. To, J.~Soto, and V.~Markl.
\newblock {A Survey of State Management in Big Data Processing Systems}.
\newblock {\em CoRR}, abs/1702.0, 2017.

\bibitem{VanSandt2019}
P.~{Van Sandt}, Y.~Chronis, and J.~M. Patel.
\newblock {Efficiently Searching In-Memory Sorted Arrays: Revenge of the
  Interpolation Search?}
\newblock In {\em Proceedings of the ACM SIGMOD International Conference on
  Management of Data}, pages 36--53, 2019.

\bibitem{Whittaker2019}
Z.~Whittaker and N.~Lomas.
\newblock {Even years later, Twitter doesn't delete your direct messages}.
\newblock {\em https://techcrunch.com/2019/02/15/twitter-direct-messages/},
  2019.

\bibitem{WiredTiger}
WiredTiger.
\newblock {Source Code}.
\newblock {\em https://github.com/wiredtiger/wiredtiger}.

\bibitem{Zhang2018a}
H.~Zhang, H.~Lim, V.~Leis, D.~G. Andersen, M.~Kaminsky, K.~Keeton, and
  A.~Pavlo.
\newblock {SuRF: Practical Range Query Filtering with Fast Succinct Tries}.
\newblock In {\em Proceedings of the ACM SIGMOD International Conference on
  Management of Data}, pages 323--336, 2018.

\bibitem{Zukowski2012}
M.~Zukowski and P.~A. Boncz.
\newblock {Vectorwise: Beyond Column Stores}.
\newblock {\em IEEE Data Engineering Bulletin}, 35(1):21--27, 2012.

\end{thebibliography}
